\documentclass[aps,prl,twocolumn,letterpaper,superscriptaddress,amsmath,amsfonts,showpacs,floatfix,longbibliography]{revtex4-2}
\usepackage{caption}
\usepackage{subcaption}
\usepackage{graphicx}
\usepackage{placeins}
\usepackage{hyperref}
\usepackage{xcolor}
\usepackage{comment}
\usepackage{listings}

\newcommand{\D}{\mathrm d}

\newcommand{\I}{\mathrm i}
\renewcommand{\vec }{\mathbf }
\newcommand{\pD}{\partial }
\newcommand{\red}{\color{red} }

\definecolor{rred}{rgb}{0.8, 0.0, 0.0}

\begin{document}

\title{Nonlocal gravity wave turbulence in presence of condensate.}

\date{\today}

\author{A.\,O.~Korotkevich}
\email{alexkor@math.unm.edu}
\affiliation{Department of Mathematics and Statistics, University of New Mexico, MSC01 1115, 1 University of New Mexico, Albuquerque, NM 87131-0001, USA}
\affiliation{L.\,D.~Landau Institute for Theoretical Physics RAS, Prosp. Akademika Semenova 1A, Chernogolovka, Moscow region, 142432, Russian Federation}

\author{S.\,V.~Nazarenko}
\email{Sergey.Nazarenko@univ-cotedazur.fr}
\affiliation{Institut de Physique de Nice, Universit\'e C\^ote d'Azur CNRS-UMR 7010,
Parc Valrose, 06108 Nice, France}

\author{Y.~Pan}
\email{yulinpan@umich.edu}
\affiliation{Department of Naval Architecture and Marine Engineering, University of Michigan, Ann Arbor, MI 48109, USA}

\author{J.~Shatah}
\email{shatah@cims.nyu.edu}
\affiliation{Courant Institute of Mathematical Sciences, New York University, 251 Mercer St., New York, NY 10012, USA}


\begin{abstract}\noindent 
We develop a theory of turbulence of weak random gravity waves on surface of deep water in which the main nonlinear process at high-frequency part of the spectrum is a nonlocal interaction with a strong low-frequency component. The latter component, which we call ``condensate", may appear in the system due to, e.g., the finite size effects which lead to an energy stagnation at waves whose wavelength is comparable to the size of the retaining flume. Our theory assumes the form of a linear spectral diffusion equation.   
We find a scaling solution of this equation and propose it as a possible explanation of recent numerical results for the gravity wave spectrum.
\end{abstract}

\maketitle
\section{Introduction.}

Wave Turbulence theory (WTT) is a statistical closure that provides a coarse-grained description for evolution of random weak wave fields \cite{ZLF1992,Nazarenko2011,galtier_2022}.
WTT has been successfully applied to very different physical cases, from quantum \cite{PROMENT2012304,quantumWT} to classical \cite{ZF1967,Zakharov1999,CONNAUGHTON20151} to cosmological \cite{NazarenkoGW} wave systems. One of the most important applications of WTT is the water surface gravity waves, understanding which is important for sea navigation and industrial offshore activities. A  review of 
most important achievements in this area can be found in \cite{NL2016}.
{ For weakly nonlinear gravity waves there are two quadratic ( in the wave amplitude) invariants:  the energy and the wave action. This system is characterised by a dual cascade behaviour, for which there exist two local Kolmogorov-Zakharov (KZ) spectra, one in which the energy is cascading downscale \cite{ZF1967} and the other one where the wave action is cascading upscale 
\cite{ZZ1982}.   Realizability of the KZ spectra require locality of interactions in the scale space which is, in turn, possible only if there are sufficiently wide inertial ranges of scales without local energy pileup in the spectral space.   However, often there is no efficient dissipation mechanism at the large scales, which leads to a spectrum pileup (or condensation) at the scales of the order of the largest available scale in the system - the size of the containing basin. Such condensation can lead to breakdown of locality of interactions. In the present Letter, we will develop a WTT description for the gravity waves in presence of a strong large-scale condensate. This will allow us to propose an explanation for deviations from the inverse-cascade KZ spectrum which were recently observed numerically in 
\cite{Korotkevich2022arXiv} and earlier reported experimentally in \cite{NL2016}.

A standard approach in describing the gravity waves  (including that
in~\cite{Korotkevich2022arXiv}) is to consider a 2D surface over a 3D potential  flow of an ideal incompressible infinitely deep fluid
with velocity $\vec v = \vec \nabla\Phi(x,y,z;t)$, where $\Phi(x,y,z;t)$ is the velocity potential. Capillary effects are neglected assuming that they are small with respect to  gravity.
%with periodic boundary conditions along the surface,  
The Hamiltonian variables for this system 
are elevation of the surface $\eta(x,y;t)$ and velocity potential on the surface $\psi(x,y;t)=\Phi(x,y,z;t)|_{z=\eta}$~\cite{Zakharov1967}.
%and these are real valued functions, 
It is convenient to introduce  normal  canonical  variables $a_{\vec k}$, corresponding to expansion in Fourier harmonics as follows,
\begin{align}
\label{a_k_substitution}
a_{\vec k} &= \sqrt \frac{\omega_k}{2k} \eta_{\vec k} + \I \sqrt \frac{k}{2\omega_k} \psi_{\vec k},\; \mbox{where   } \omega_k = \sqrt {g k}.
\end{align}
In terms of these variables, the  surface dynamics is Hamiltonian
\begin{align}
\label{Hamiltonian_eqs_canonical}
\dot a_{\vec k} &= -\I \frac{\delta H}{\delta a_{\vec k}^{*}}
\end{align}
with the Hamiltonian $H$ being the total physical energy of the wave system
(not shown).

For statistical description of a stochastic wave field one can use a pair correlation function called the wave action spectrum,
\begin{equation*}
\langle a_{\vec k} a_{\vec k'}^*\rangle = n_{\vec k} \delta (\vec k - \vec k').
\end{equation*}
The spectrum $n_{\vec k}$ is a measurable quantity, directly related to observable correlation functions, e.g., 
from the  definition of $a_{\vec k}$, one can get
\begin{equation*}
\label{I_k_expression}
I_{\vec k} = \langle |\eta_{\vec k}|^2\rangle = \frac{1}{2}\frac{\omega_k}{g} (n_{\vec k} + n_{-\vec k}).
\end{equation*}
\begin{comment}
To derive kinetic equation in the case of gravity waves it is convenient to use another correlation function
\begin{equation*}
\langle b_{\vec k} b_{\vec k'}^*\rangle = N_{\vec k} \delta (\vec k - \vec k'),
\end{equation*}
where $b_{\vec k}$ is an analog of $a_{\vec k}$ obtained as a result of canonical transformation eliminating non-resonant three-wave interaction terms in the Hamiltonian. For the cases we are interested in the relative difference between $N_k$ and $n_k$ is at most few percents for average steepness of the surface $\mu=\sqrt{\langle|\vec\nabla\eta|^2\rangle}\approx 0.1$. So we shall use $n_k$ for analysis of numerical results and neglect difference between $N_{\vec k}$ and $n_{\vec k}$.
\end{comment}
}
Under the WTT assumptions, i.e. small wave amplitudes, random phases and large-basin limit, spectrum $n_k$ obeys the Hasselmann wave-kinetic equation (WKE)~\cite{Nordheim1928,Peierls1929,Hasselmann1962}:
\begin{equation}
\label{Kinetic_equation}
\frac{\partial n_{\vec k}}{\partial t} = I_{St} + f_p (k) - f_d (k),
\end{equation}
Here $f_p$ and $f_d$ are some pumping and damping terms respectively, and $I_{St}$ is a so called collision integral:
\begin{align}
I_{St}&=4\pi \int \left| T_{\vec k,\vec k_1}^{\vec k_2,\vec k_3}\right|^2 n_{\vec k} n_{\vec k_1} n_{\vec k_2} n_{\vec k_3}\left(\frac{1}{n_{\vec k}} + \frac{1}{n_{\vec k_1}} -\right.\nonumber\\
&\left. - \frac{1}{n_{\vec k_2}} - \frac{1}{n_{\vec k_3}}\right)\delta (\vec k + \vec k_1- \vec k_2 - \vec k_3)\times\label{collision_integral}\\
&\delta (\omega_k + \omega_{k_1}- \omega_{k_2} - \omega_{k_3}) \D \vec k_1\D \vec k_2\D \vec k_3.\nonumber
\end{align}
The interaction coefficients $T_{\vec k,\vec k_1}^{\vec k_2,\vec k_3}$ can e.\,g. be found in~\cite{PRZ2003} and in Chapter II of the Supplemental material.
The kinetic equation and its modifications are the basis for all wave forecasting models.

One can consider stationary solutions of WKE in the so called inertial interval -- range of scales far from pumping and damping regions. Such solutions have to obey $I_{St}=0$ equation. Beyond obvious Raleigh-Jeans thermodynamic equilibrium (zero flux) solutions, $n_{\vec k} \sim 1/(\omega_k +$~const), there are dynamic equilibrium solutions, corresponding to  finite fluxes of conserved quantities, the so-called KZ spectra~\cite{ZLF1992,Nazarenko2011}.  Equations~\eqref{Kinetic_equation}-\eqref{collision_integral} describe a four-wave process of scattering of two waves into  two waves. This means that, in addition to the total energy $E=\int \omega_k n_{\vec k} \D\vec k$, there is a conservation of
the total wave action  (``number of waves'') $N=\int n_{\vec k} \D\vec k$.
{\red Thus, }there are two KZ spectra: one describing a local downscale (with respect to the forcing scale) energy cascade,
\begin{equation}
n_k^{DC} = C_P P^{1/3} k^{-4},\label{WTT_direct_spectrum}
\end{equation}
and the other one - a local upscale (toward smaller $\vec k$'s) wave action cascade~\cite{ZZ1982},
\begin{equation}
n_k^{IC} = C_Q Q^{1/3} k^{-23/6}\sim k^{-3.83}\label{WTT_inverse_spectrum}.
\end{equation}
Here, $P$ and $Q$ are the fluxes of energy and wave action respectively, $C_P$ and $C_Q$ are some constants, and $k=|\vec k|$. 

In a recent paper~\cite{Korotkevich2022arXiv}, the primordial dynamical equations were simulated in a region $L_x = L_y = 2\pi$ with double periodic boundary conditions, which means that $k$'s were integer numbers. Grid resolution was $N_x=N_y=512$. Pumping was isotropic with respect to angle and concentrated in a ring $k \in (60;64)$ with random phase of every pumped harmonic. Damping started at $k_d=128$. As a result, in the range of wavenumbers smaller than $k=60$ one would expect to find a spectrum similar to~\eqref{WTT_inverse_spectrum}. Nevertheless, it was reported, that the observed inverse cascade spectrum had a different slope, close to $n_k \sim k^{-3.07}$, which was indistinguishable for significantly different levels of nonlinearity in the system~\cite{Korotkevich2022arXiv}. This numerically observed spectrum is rather close to previous experimental results which showed $n_k \sim k^{-2.5}$ or $n_k \sim k^{-3}$ depending on the forcing frequency~\cite{NL2016}.

\begin{figure}[htb!]
\centering
    \begin{subfigure}[t]{0.25\textwidth}
        \centering
        \includegraphics[width=\linewidth]{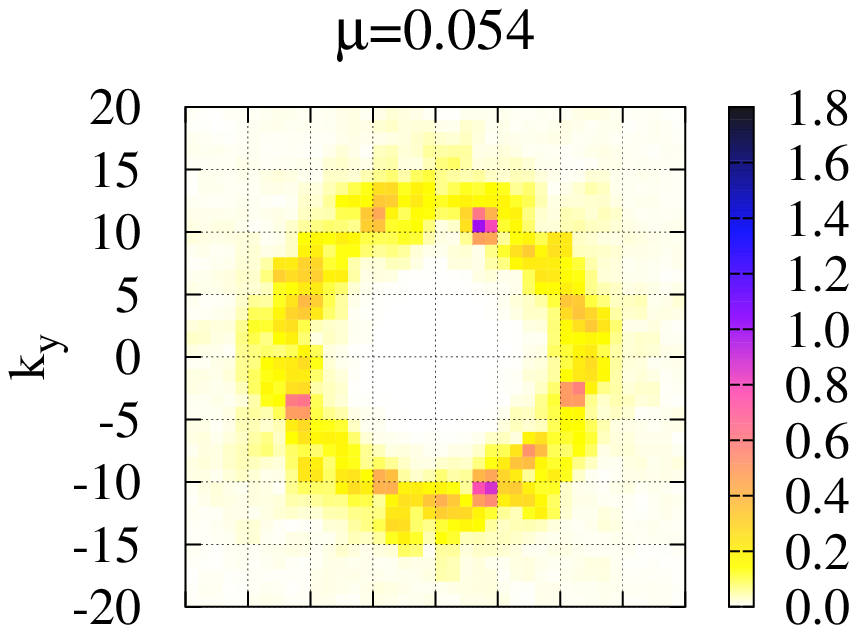}
    \end{subfigure}
    %\hfill
    \begin{subfigure}[t]{0.20\textwidth}
        \centering
        \includegraphics[width=\linewidth]{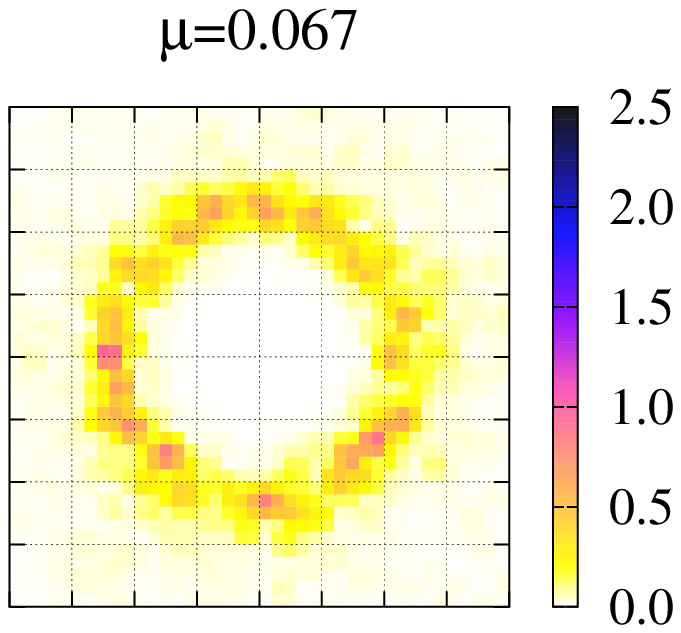}
    \end{subfigure}

    %\vspace{1cm}
    \begin{subfigure}[t]{0.25\textwidth}
    \centering
        \includegraphics[width=\linewidth]{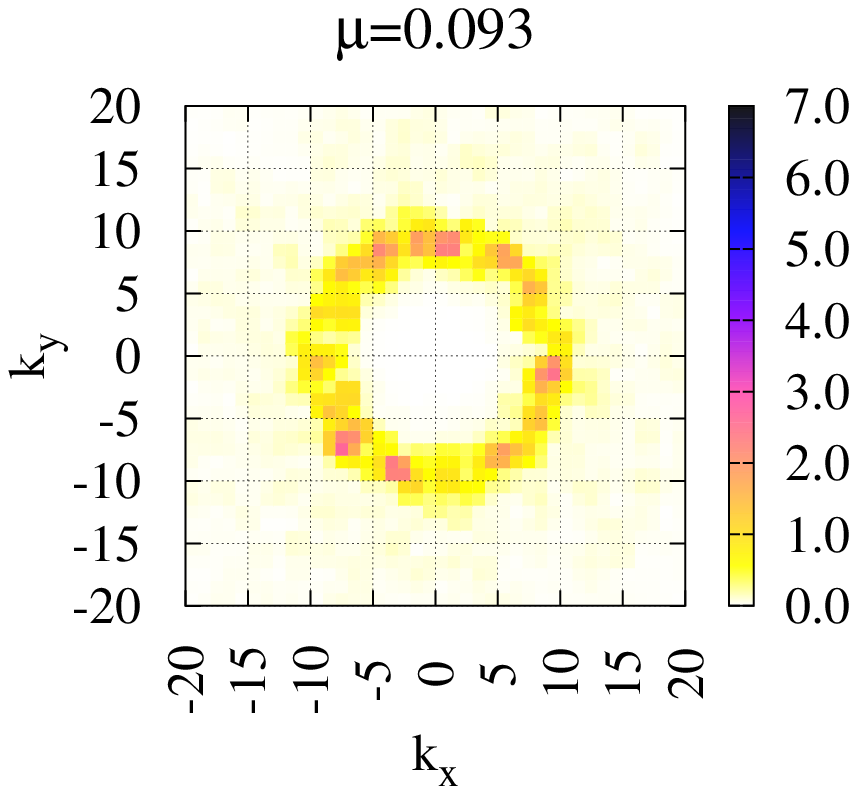}
    \end{subfigure}
    %\hfill
    \begin{subfigure}[t]{0.20\textwidth}
        \centering
        \includegraphics[width=\linewidth]{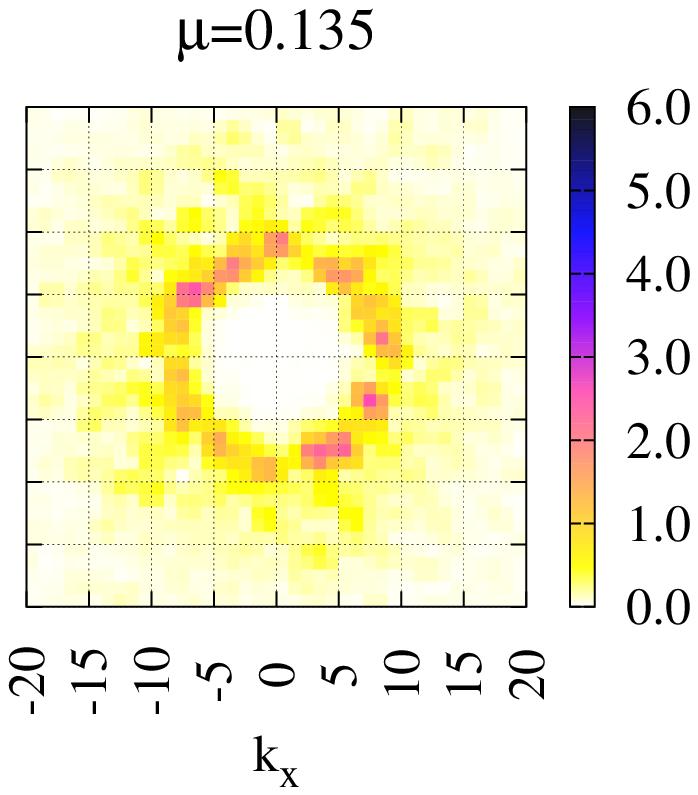}
    \end{subfigure}
    \caption{\label{condensate_rings} Surfaces of $|a_{\vec k}|^2\times 10^{8}$ in the condensate region for all simulation  in~\cite{Korotkevich2022arXiv}.  $\mu$ is the average wave slope.}
\end{figure}

In previous works~\cite{Korotkevich2008PRL,Korotkevich2012MCS} it was shown that condensate plays an important role in the nonlinear interaction processes for such systems. Specifically, the measured dispersion relation~\cite{Korotkevich2013JETPL} demonstrated that influence of the condensate on the harmonics in the region of inverse cascade cannot be neglected. There are several approaches which allow one to take the condensate influence into account. Probably the most obvious one is based on Bogolyubov's transformation technique, like in the case
of dilute Bose gas,  similar to the recent work in~\cite{GKLN2022} (see sections about degenerated almost ideal Bose gas in~\cite{AGD1962} or~\cite{LL_vol9}). This approach requires the condensate to be a coherent object, which is usually achieved due to the fact that it is located at a single harmonic, $k=0$. In the case of numerical simulations in~\cite{Korotkevich2022arXiv}, the condensate was roughly a ring in the $k$-space, whose further cascade to low wavenumbers is prohibited by the finite size effect of the domain. The pictures of the $|a_{\vec k}|^2$-distributions are shown in Figure~\ref{condensate_rings}.
One can see that the wave amplitudes are randomized and, in most cases, are significantly different from each other in amplitude even for harmonics close in $\bf k$. So the Bogolyubov approach is clearly not applicable. There are more than a hundred harmonics in the condensate ring. Since  averaging over an ensemble of realizations is  computationally unfeasible, we average the $|a_{\vec k}|^2$-function over the azimuthal angle  and obtain an object $\langle |a_{\vec k}|^2\rangle$ which can be considered as an approximation for $n_k$, 
Thus, following ideas proposed in~\cite{Zakharov2010}, one can try to modify the WKE~\eqref{Kinetic_equation} to describe the statistics of the wave field on the background of condensate.

The key development, as we will discuss in the next section, is the derivation of a diffusion equation from the wave kinetic equation of surface gravity waves in presence of condensate. We put an alternative derivation in Chapter III of the Supplemental Material, which relies on a WKB approximation of the Zakharov equation. The two approaches yield exactly the same result including the detailed formulation of the diffusion coefficient. Such diffusion equations were mainly developed for three-wave-resonant systems in the past, such as internal gravity waves~\cite{MCC1977}, Rossby waves~\cite{CONNAUGHTON20151} and MHD turbulence~\cite{NAZ2001}. In addition to establishing the diffusion mechanism for surface gravity waves as a four-wave-resonant system, we also show that the constant-flux stationary solution of the diffusion equation yields $n_k \sim k^{-3}$, serving as an explanation for the previous numerical and experimental results.  

\section{Derivation of diffusion equation in presence of condensate.}
\begin{figure}
\centering
\includegraphics[width=3.5in]{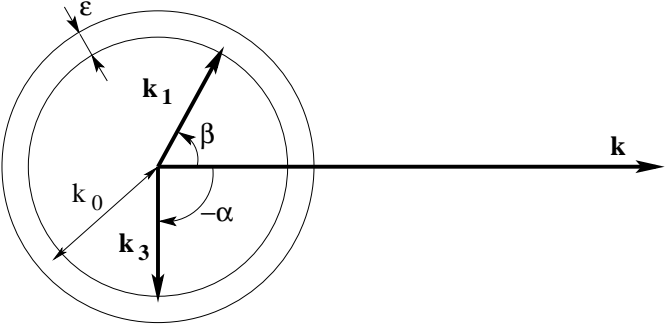}
\caption{\label{fig:condensate_scheme}Scheme of considered wave vectors with respect to position of condensate ring.}
\end{figure}

%Let us look for a constant flux solution for wave action $n_k$ of WKE~\eqref{Kinetic_equation}.
Assume that the condensate modes are much stronger than any other harmonics, so that  for any four waves in the resonant quartet, the dominant contribution comes from the situation that two waves $\vec k_1$ and $\vec k_3$ are in the condensate (see Figure~\ref{fig:condensate_scheme}), satisfying
\begin{equation}
\vec k + \vec k_1 = \vec k_2 + \vec k_3,\;\; \omega_k + \omega_{k_1} = \omega_{k_2} + \omega_{k_3}.\label{resonance_conditions}
\end{equation}
We consider the case $k\gg k_{1,3}$ and strong condensate $n_{\vec k_{1,3}}\gg n_{\vec k}, n_{\vec k_2}$. The case when both condensate waves are at one side of resonant conditions~\eqref{resonance_conditions} cannot be realized.
Situation is isotropic with respect to azimuthal angle. For simplicity let us consider
the case where the condensate is supported  in a ring as in Figure~\ref{fig:condensate_scheme}. 
Writing $\vec q = \vec k_1 - \vec k_3$ for the difference of wavevectors, i.e.,  $\vec k_2 = \vec k + \vec q$, and 
%$k\gg k_0$, so $k_0/k$ is a small parameter. 
neglecting $1/n_{\vec k_{1,3}}$ terms in~\eqref{collision_integral} one gets:
\begin{equation}
\label{first_difference}
\begin{array}{l}
\displaystyle
\frac{\partial n_{\vec k}}{\partial t} \approx 8\pi
%\int \left| T_{\vec k,\vec k_1}^{\vec k+\vec q,\vec k_3}\right|^2 n_{\vec k} n_{\vec k_1} n_{\vec k+\vec q} n_{\vec k_3}\times\\ \displaystyle\left(\frac{1}{n_{\vec k}} - \frac{1}{n_{\vec k+\vec q}}\right) \delta (\omega_k + \omega_{k_1}- \omega_{k_2} - \omega_{k_3}) \D \vec k_1\D \vec k_3=\\ \displaystyle
 \int \Big\{\left| T_{\vec k,\vec k_1}^{\vec k_2,\vec k_3}\right|^2 n_{\vec k_1} n_{\vec k_3}( n_{\vec k_2} - n_{\vec k})\\
\displaystyle
\times
\delta (\varOmega)\Big\}_{\vec k_2 = \vec k+\vec q}
\D \vec k_1\D \vec k_3,
\end{array}
\end{equation}
where $\varOmega = \omega_k + \omega_{k_1}- \omega_{k_2} - \omega_{k_3}$. The additional factor of two relative to~\eqref{collision_integral} is due to the fact that the condensate is supported either on $\vec k_2$ or $\vec k_3$.

Next we exploit Taylor expansion in $\vec q$ of the expression 
\[
 \left|T_{\vec k,\vec k_1}^{\vec k_2,\vec k_3}\right|^2( n_{\vec k_2} - n_{\vec k}),
 \]
here $\vec k_2= \vec k + \vec q$, to the second order (see Chapter I in Supplemental Material):
\begin{equation}
\label{second_difference}
\begin{array}{l}
\displaystyle
\frac{\partial n_{\vec k}}{\partial t} \approx 8\pi
 \int n_{\vec k_1} n_{\vec k_3} \left|T_{\vec k,\vec k_1}^{\vec k, \vec k_3}\right|^2\vec q\cdot\vec\nabla_{\vec k}n_{\vec k}
\left.\delta (\varOmega)\right|_{\vec k_2 = \vec k+\vec q}
\D \vec k_1\D \vec k_3+\\
\displaystyle
+4\pi\int n_{\vec k_1} n_{\vec k_3} \left\{ 2\left.\left(\vec q\cdot\vec\nabla_{\vec k_2}\left|T_{\vec k,\vec k_1}^{\vec k_2, \vec k_3}\right|^2\right)\right|_{\vec k_2 =\vec k}\vec q\cdot\vec\nabla_{\vec k}n_{\vec k} +\right.\\
+ \left.\left|T_{\vec k,\vec k_1}^{\vec k, \vec k_3}\right|^2\vec q\cdot\vec\nabla_{\vec k}\left(\vec q\cdot\vec\nabla_{\vec k}n_{\vec k}\right)
\right\}\left.\delta (\varOmega)\right|_{\vec k_2 = \vec k+\vec q}
\D \vec k_1\D \vec k_3,
\end{array}
\end{equation}

To simplify \eqref{second_difference}, we  consider the expansion of $\varOmega$ in $\vec q$:
\[
\begin{split}
\varOmega = \; & \omega_{k_1} -\omega_{k_3} -\nabla_{\vec k}\omega_k\cdot \vec q + \mathcal{O}(|\vec q|^2).
\end{split}
\]
Using the equation $\omega_k= \sqrt{gk}$, we obtain
\begin{flalign}
\varOmega \approx& \; \omega_{k_1}-\omega_{k_3} -\frac{\sqrt{g}}{2} \frac{\vec k}{k^{3/2}}\cdot(\vec k_1- \vec k_3)=\nonumber\\
= &  \; \omega_{k_1}-\omega_{k_3} -\frac{\sqrt{g}}{2}\frac{k_1\cos\alpha - k_3\cos\beta}{k^{1/2}},\label{approxOmega}
\end{flalign}
where $\alpha$ and $\beta$ are in Figure~\ref{fig:condensate_scheme}.
Here we keep the next order term which becomes part of the leading one if $k_1\approx k_3$.
It should be noted that $\delta(\varOmega)$ (here and further we use the last expression for $\varOmega$) is invariant with respect to exchange $\vec k_1\leftrightarrow \vec k_3$. Using this symmetry as well as $T_{\vec k,\vec k_1}^{\vec k_2, \vec k_3}=T_{\vec k_2,\vec k_3}^{\vec k, \vec k_1}$ and taking into account a fact that $\vec k_1$ and $\vec k_3$ are dummy variables so that value of an integral must not change after exchange $\vec k_1 \leftrightarrow \vec k_3$, the first term of \eqref{second_difference} is integrated to zero, due to change of sign of $\vec q$ after the mentioned exchange of integration variables. 
 For evaluation of the second term in~\eqref{second_difference} at $\vec k_2=\vec k$ one can use the following equality:
\begin{equation}
\!\!\left.\vec\nabla_{\vec k_2}\left|T_{\vec k,\vec k_1}^{\vec k_2, \vec k_3}\right|^2\right|_{\vec k_2 =\vec k} \!\!\!\!\!\!\!\!= \left.\vec\nabla_{\vec k}\left|T_{\vec k_2,\vec k_1}^{\vec k, \vec k_3}\right|^2\right|_{\vec k_2 =\vec k} \!\!\!\!\!\!= \frac{1}{2} \vec\nabla_{\vec k}\left|T_{\vec k,\vec k_1}^{\vec k, \vec k_3}\right|^2\!\!\!\!.
\end{equation}
where the second equality should be understood in the context of integration over $\vec k_1$ and $\vec k_3$ as in \eqref{second_difference}. As a result~\eqref{second_difference} can be transformed (after combining the second and third terms) into the following expression:
\begin{equation}
\label{third_difference}
\begin{split}
\displaystyle
\frac{\partial n_{\vec k}}{\partial t} \approx& 
 4\pi\int n_{\vec k_1} n_{\vec k_3}
 \vec\nabla_{\vec k}\cdot\left(\vec q \left|T_{\vec k,\vec k_1}^{\vec k, \vec k_3}\right|^2\vec q\cdot\vec\nabla_{\vec k}n_{\vec k}\right)\\
\displaystyle
&\times\delta (\varOmega)
\D \vec k_1\D \vec k_3 .
%= \\\displaystyle
\end{split}
\end{equation}
This is a continuity equation for the wave action $n_{\vec k}$,
\begin{equation}
\frac{\partial n_{\vec k}}{\partial t} + \vec \nabla_{\vec k}\cdot\vec Q_{\vec k}=0.
\end{equation}
with action flux $\vec Q_{\vec k}$ for an isotropic spectrum, where 
$\vec Q_{\vec k} = - 4\pi \int n_{\vec k_1}n_{\vec k_3} \vec q \left| T_{\vec k,\vec k_1}^{\vec k,\vec k_3}\right|^2
\frac{\vec q\cdot\vec k}{k} \frac{\pD n_{k}}{\pD k}
\delta (\varOmega)
\D \vec k_1\D \vec k_3$.  Under
 the same isotropic consideration we have
\begin{align}
N=\int n_{\vec k}\D\vec k &=\int n_{k}
2\pi k\D k,
%\;\; \vec Q_{\vec k} = \frac{\vec k}{k} Q_k,
\\
\frac{\partial (2\pi k n_k)}{\partial t} &+\frac{\partial  Q_{ k}}{\partial k}=0.
%\nonumber
\end{align}
Here we introduced the isotropic wave action flux
\begin{eqnarray}
Q_k =2\pi\vec k\cdot\vec Q_{\vec k}= -\left[\frac {8\pi^2} k \int n_{\vec k_1}n_{\vec k_3}
%n_{\vec k_3}
\delta (\varOmega)\times\nonumber\right.\\
\left.\times\left(\left| T_{\vec k,\vec k_1}^{\vec k,\vec k_3}\right|^2 (\vec q\cdot\vec k)^2 \right)
\D \vec k_1\D \vec k_3\right]\frac{\partial n_{k}}{\partial k}.
\label{eqQ}
\end{eqnarray}
One can integrate out $k_3$ using $\delta(\varOmega)$. In order to do this we need to set $\varOmega=0$ in~\eqref{approxOmega}. Solving this quadratic equation for $\sqrt{k_3}$ and taking into account condition $k_1\ll k$ we get:
\begin{flalign}
k_3 \approx k_1 -k_1\sqrt{\frac{k_1}{k}}(\cos\alpha-\cos\beta).
\label{k1k3}
\end{flalign}
The integration in~\eqref{eqQ} is over $0\le \alpha,\beta <2\pi$ and $k_1$. If the relative width of the  condensate ring $\varepsilon/k_0$ is small, then from \eqref{k1k3} $k_3$ will be outside the ring for various values of $\alpha, \beta \in (0,2\pi)$, see Figure~\ref{fig:condensate_scheme}. However if the condensate ring is wide $\varepsilon/k_0 \gg  2 (k_0/k)^{1/2}$ we can always find $k_3 \approx k_1$ for all $0\le \alpha,\beta <2\pi$ (details can be  found in Chapter~IV of Supplemental Materials).  As a result of a strong inequality we can  take $n_{k_3}=n_{k_1}$, thus we have a diffusion equation
\begin{equation}
\label{diffusion_equation}
2\pi k \frac{\partial n_{k}}{\partial t} = \frac{\partial}{\partial k} \left(D_k \frac{\partial n_{k}}{\partial k} \right),
\end{equation}
with  diffusion coefficient 
\begin{align}
\label{diffusion_coeff}
&D_k = 8\pi^2 k \int\limits_{k_1 \ll k}n_{k_1}^2\frac{k_1^4}{\left|\frac{d \omega_{k_1}}{d k_1}\right|}\times\nonumber\\
&\left[\int\limits_{0}^{2\pi}
\int\limits_{0}^{2\pi} 
\left| T_{\vec k,\vec k_1}^{\vec k,\vec k_3}\right|^2 (\cos\beta-\cos\alpha)^2
\D\alpha\D \beta\right]\D k_1.
\end{align}

In Chapter II of  Supplemental Material it is shown that the first non-vanishing term in the expansion of $T_{\vec k,\vec k_1}^{\vec k,\vec k_3}$ in    $k_{1,3}/k$  (after  cancellation of all terms of order $k^2k_1$) is, 
\begin{equation}
\label{T_approx}
%T_{\vec k_1\vec k}^{ \vec k_3\vec k} 
T_{\vec k,\vec k_1}^{\vec k,\vec k_3}= -\frac{(k k_1)^{3/2}}{16\pi^2}(\cos\alpha -\cos\beta)^2.
\end{equation}
To find the diffusion coefficient~\eqref{diffusion_coeff} one needs to compute the following integral using~\eqref{T_approx},
$$
\int\limits_{0}^{2\pi}\int\limits_{0}^{2\pi} \left| T_{\vec k,\vec k_1}^{\vec k,\vec k_3}\right|^2 (\cos\beta-\cos\alpha)^2
\D\alpha\D \beta = \frac{25(k k_1)^{3}}{256\pi^2},
$$
which results in
\begin{equation}
\label{diffusion_coeff_approx}
  D_k %= \pi k \int\limits_{k_1 \ll k}n_{k_1}^2 k_1^{9/2} \frac{25(k k_1)^{3}}{256\pi^2}\D k_1
%= \frac{25}{256\pi} k^4 \int\limits_{k_1 \ll k}n_{k_1}^2 k_1^{15/2} \D k_1
=\lambda k^4, \; \lambda =  \frac{25\pi}{16} \int\limits_{k_1 \ll k}n_{k_1}^2 k_1^{15/2} \D k_1 = \hbox{const} .
\end{equation}
%\textcolor{blue}{Can we compute $D$ for small $\epsilon$?}
Note also, that we lose the universal (i.e. independent of the spectrum shape) scaling $D_k \sim k^4$ if width of the condensate $\varepsilon/k_0
\sim (k_0/k)^{1/2}$.

%From~\eqref{diffusion_coeff_approx} it follows that $D_k\sim k^4$, so if we 
 From
$Q=-D_k \pD n_{k}/\pD k = const$,  we immediately get 
%$\partial n_{k}/\partial k \sim k^{-4}$, which after integration gives $n_{k}\sim k^{-3}$.
\begin{equation}
\label{nonlocN}
n_{k} = \frac{Q}{3\lambda} k^{-3},
\end{equation}
which is a constant-flux power law solution.
Note that, since both $n_{k}$ and $\lambda$ are positive,  the wave action flux $Q$ is positive too, i.e. it is toward high $k$ and has the opposite direction with respect to the wave action flux on the local KZ solution with $Q=$~const. But by the standard Fjortoft argument the wave action cannot be continuously transferred to high $k'$s -- otherwise the amount of energy transferred to these wave numbers would be greater than the energy produced by the forcing, which is impossible. Therefore, the nonlocal cascade must drain wave action from the condensate.
Similarly, the energy is also drained from the condensate. Indeed, for the energy of the out-of-condensate modes we have from \eqref{diffusion_equation}:
\begin{equation}
\label{Edot}
\dot E= \int \omega_k (D_k n_k')' dk = \frac{7\lambda}{4} \int k^{5/2} n_k dk >0.
\end{equation}
Thus, for any shape of $n_k$ the energy of the out-of-condensate modes is growing and, since the total energy must be conserved, in absence of forcing and dissipation, we conclude that the condensate energy must be decreasing at the same rate. We also conclude that for a steady state to exist there should exist a mechanism of supplying the energy and the wave action from the forcing region to the condensate which is not the  diffusive  mechanism arising from the WKE in the nonlocal regime (since the latter transfers these invariants in the opposite direction).

\section{Conclusions and discussion.}

In this Letter, we considered turbulence of water surface gravity waves and developed a WTT for nonlocal wave action spectrum evolution in presence of a strong large-scale condensate. We derived a singular inhomogeneous spectral diffusion equation governing such an evolution. We found a stationary solution corresponding to a constant wave action cascade, $n_{k}\sim k^{-3}$, and proposed it may explain the spectrum observed numerically in \cite{Korotkevich2022arXiv} and experimentally in \cite{NL2016}.

Regarding the mechanism of supply of energy and wave action to the condensate, one can envisage two scenarios. First, the condensate could be built up via a local inverse cascade at the initial stages, followed by the nonlocal regime switching the cascade direction. Clearly, in this case the nonlocal regime would not be able to sustain itself due to the condensate drainage and the system would return to the local regime, with a possible periodic repetition of the local-nonlocal cycle. We rule out this possibility because no such oscillations are observed in the numerical simulations. The second possibility is excitation of the condensate modes directly from quasi-resonant three-wave process following modulational instability of the forcing modes. Such a mechanism requires that the spectrum of the waves in the forcing range is sufficiently strong --  clearly a feature seen in the numerical results of  \cite{Korotkevich2022arXiv}. Moreover, the latter process could remain active simultaneously with the nonlocal direct cascade and thereby contribute to formation of a stationary spectrum. However, this scenario remains speculative and requires further validation.

\begin{acknowledgments}
The authors are grateful for support from the Simons' Collaboration on Wave Turbulence (awards \#651459 and \#651459). The research was partially performed during KAO's visit to the the Universit\'e C\^ote d'Azur/Institut de Physique de Nice, France, funded by F\'ed\'eration de Recherche ``Wolfgang D\"oblin'' and ``Waves Complexity'' visiting researcher program, and visits of KAO, NSV, and PY to the Courant Institute.
\end{acknowledgments}

\onecolumngrid
\appendix

\section{Supplemental Material: Description. }
The four sections of the supplemental materials describe respectively Taylor expansion up to the second order leading to continuity equation for wave action, reduction of interaction coefficient, the derivation of spectral diffusion equation from the Zakharov equation for surface gravity waves, and consideration of the range of applicability of the proposed theory based on different parameters.

\section{Supplemental Material: Part I. Taylor expansion of an integrand up to the second order.}
Let us start from the $\vec k$ and $\vec k_2$ dependent part of an integrand. We also will use the fact that $\vec k_1$ and $\vec k_3$ are dummy variables (used for integration) and can be exchanged with proper adjustment of an integrand without the change of the value of an integral. We shall consider Taylor expansion of the following expression:
$$
F=\left|T_{\vec k,\vec k_1}^{\vec k_2, \vec k_3}\right|^2(n_{\vec k_2} - n_{\vec k}),
$$
with respect to $\vec k_2 = \vec k + \vec q$ around $\vec k$, here $\vec q = \vec k_1 - \vec k_3$. It should be noted that $T_{\vec k,\vec k_1}^{\vec k_2, \vec k_3}$ is a homogeneous function, so one could divide all $\vec k_i$'s by $k$ and use the fact that we consider the setting where $k_{1,3}/k\ll 1$, so $|\vec q|/k\ll 1$ and expansion of $T_{\vec k,\vec k_1}^{\vec k_2, \vec k_3}$ considering $|\vec q|$ small with respect to $k$ is a valid procedure. Also we suppose that function $n_{\vec k}$ has all necessary properties, so it can be expanded in terms of $\vec q$ considering it as a small perturbation. For instance, numerical data in~\cite{Korotkevich2022arXiv} supposed $n_{k}\sim k^{-3.07}$ which is decaying power-law function which again allows us to divide all $\vec k_i$'s by $k$, so again Taylor expansion assuming $|\vec q|$ small with respect to $k$ is a reasonable procedure.

The zeroth order of the expansion is obviously zero:
$$
F|_{\vec k_2 =\vec k} = 0.
$$
The first order of the expansion is (before evaluation at $\vec k_2=\vec k$):
\begin{equation}
\label{Taylor_first_order}
D_{\vec k_2} F = \left(\vec q\cdot\vec\nabla_{\vec k_2}\left|T_{\vec k,\vec k_1}^{\vec k_2, \vec k_3}\right|^2\right)(n_{\vec k_2} - n_{\vec k}) + \left|T_{\vec k,\vec k_1}^{\vec k_2, \vec k_3}\right|^2\vec q\cdot\vec\nabla_{\vec k_2}n_{\vec k_2}.
\end{equation}
Evaluated at $\vec k_2 =\vec k$ this expression is:
\begin{equation}
\label{Taylor_first_order_evaluated}
\left.D_{\vec k_2} F\right|_{\vec k_2 =\vec k} = \left|T_{\vec k,\vec k_1}^{\vec k, \vec k_3}\right|^2\vec q\cdot\vec\nabla_{\vec k}n_{\vec k}.
\end{equation}
Using the symmetry $T_{\vec k,\vec k_1}^{\vec k_2, \vec k_3}=T_{\vec k_2,\vec k_3}^{\vec k, \vec k_1}$ and taking into account already mentioned fact that $\vec k_1$ and $\vec k_3$ are dummy variables and value of an integral must not change after exchange $\vec k_1 \leftrightarrow \vec k_3$, this expression will be integrated to zero, because the integral of the~\eqref{Taylor_first_order_evaluated} changes sign due to change of sign of $\vec q$ after the mentioned exchange of integration variables. So one needs to consider the next order of the expansion.

Using~\eqref{Taylor_first_order}, the second order of the expansion is (before evaluation at $\vec k_2=\vec k$):
\begin{flalign}
D_{\vec k_2}^2 F &= D_{\vec k_2}(D_{\vec k_2} F) = \vec q\cdot\vec\nabla_{\vec k_2}\left(\vec q\cdot\vec\nabla_{\vec k_2}\left|T_{\vec k,\vec k_1}^{\vec k_2, \vec k_3}\right|^2\right)(n_{\vec k_2} - n_{\vec k}) + \left(\vec q\cdot\vec\nabla_{\vec k_2}\left|T_{\vec k,\vec k_1}^{\vec k_2, \vec k_3}\right|^2\right)\vec q\cdot\vec\nabla_{\vec k_2}n_{\vec k_2} +\nonumber\\
&+\left(\vec q\cdot\vec\nabla_{\vec k_2}\left|T_{\vec k,\vec k_1}^{\vec k_2, \vec k_3}\right|^2\right)\vec q\cdot\vec\nabla_{\vec k_2}n_{\vec k_2} + 
|T_{\vec k,\vec k_1}^{\vec k_2, \vec k_3}|^2\vec q\cdot\vec\nabla_{\vec k_2}\left(\vec q\cdot\vec\nabla_{\vec k_2}n_{\vec k_2}\right)=\nonumber\\
&=\vec q\cdot\vec\nabla_{\vec k_2}\left(\vec q\cdot\vec\nabla_{\vec k_2}\left|T_{\vec k,\vec k_1}^{\vec k_2, \vec k_3}\right|^2\right)(n_{\vec k_2} - n_{\vec k}) +\nonumber\\
&+ 2\left(\vec q\cdot\vec\nabla_{\vec k_2}\left|T_{\vec k,\vec k_1}^{\vec k_2, \vec k_3}\right|^2\right)\vec q\cdot\vec\nabla_{\vec k_2}n_{\vec k_2} + \left|T_{\vec k,\vec k_1}^{\vec k_2, \vec k_3}\right|^2\vec q\cdot\vec\nabla_{\vec k_2}\left(\vec q\cdot\vec\nabla_{\vec k_2}n_{\vec k_2}\right).\label{Taylor_second_order}
\end{flalign}
Evaluating~\eqref{Taylor_second_order} at $\vec k_2=\vec k$ one gets:
\begin{flalign}
\left.D_{\vec k_2}^2 F\right|_{\vec k_2 =\vec k} = 2\left.\left(\vec q\cdot\vec\nabla_{\vec k_2}\left|T_{\vec k,\vec k_1}^{\vec k_2, \vec k_3}\right|^2\right)\right|_{\vec k_2 =\vec k}\vec q\cdot\vec\nabla_{\vec k}n_{\vec k} + \left|T_{\vec k,\vec k_1}^{\vec k, \vec k_3}\right|^2\vec q\cdot\vec\nabla_{\vec k}\left(\vec q\cdot\vec\nabla_{\vec k}n_{\vec k}\right).\label{Taylor_second_order_evaluated_1}
\end{flalign}
Using again the symmetry $T_{\vec k,\vec k_1}^{\vec k_2, \vec k_3}=T_{\vec k_2,\vec k_3}^{\vec k, \vec k_1}$ and the fact that every term in~\eqref{Taylor_second_order_evaluated_1} has two factors $\vec q$, so exchange $\vec k_1 \leftrightarrow \vec k_3$ does not changes sign of terms, for evaluation of the first term of~\eqref{Taylor_second_order_evaluated_1} at $\vec k_2=\vec k$ one can use the following equality:
\begin{equation}
\left.\vec\nabla_{\vec k_2}\left|T_{\vec k,\vec k_1}^{\vec k_2, \vec k_3}\right|^2\right|_{\vec k_2 =\vec k} = \left.\vec\nabla_{\vec k}\left|T_{\vec k_2,\vec k_1}^{\vec k, \vec k_3}\right|^2\right|_{\vec k_2 =\vec k} = \frac{1}{2} \vec\nabla_{\vec k}\left|T_{\vec k,\vec k_1}^{\vec k, \vec k_3}\right|^2.
\end{equation}
As a result we get the following expansion of $F$ up to the second order:
\begin{flalign}
F=&\left|T_{\vec k,\vec k_1}^{\vec k_2, \vec k_3}\right|^2(n_{\vec k_2} - n_{\vec k})\approx \left(\vec q\cdot\vec\nabla_{\vec k}\left|T_{\vec k,\vec k_1}^{\vec k, \vec k_3}\right|^2\right)\vec q\cdot\vec\nabla_{\vec k}n_{\vec k} + \left|T_{\vec k,\vec k_1}^{\vec k, \vec k_3}\right|^2\vec q\cdot\vec\nabla_{\vec k}\left(\vec q\cdot\vec\nabla_{\vec k}n_{\vec k}\right)=\nonumber\\
=& \vec q\cdot\vec\nabla_{\vec k}\left(\left|T_{\vec k,\vec k_1}^{\vec k, \vec k_3}\right|^2\vec q\cdot\vec\nabla_{\vec k}n_{\vec k}\right)=
\vec\nabla_{\vec k}\cdot\left(\vec q \left|T_{\vec k,\vec k_1}^{\vec k, \vec k_3}\right|^2\vec q\cdot\vec\nabla_{\vec k}n_{\vec k}\right).
\end{flalign}

\section{Supplemental Material: Part II. Interaction coefficient reduction.}
We reproduce formula for the matrix element of interaction (interaction coefficient) from~\cite{Zakharov1999} with corrections from~\cite{PRZ2003} (here we introduce notation $\omega_i = \omega_{\vec k_i}$):
\begin{flalign}
&T_{\vec k_1 \vec k_2}^{\vec k_3 \vec k_4,{PRZ}} = \frac{1}{2}(\tilde T_{\vec k_1 \vec k_2}^{\vec k_3 \vec k_4} + \tilde T_{\vec k_2 \vec k_1}^{\vec k_3 \vec k_4}),\label{T_1234_general_PZ}\\
&\tilde T_{\vec k_1 \vec k_2}^{\vec k_3 \vec k_4} = -\frac{1}{16\pi^2}\frac{1}{(k_1 k_2 k_3 k_4)^{1/4}}\nonumber\\
&\times\left\{-12k_1 k_2 k_3 k_4 - 2(\omega_1 + \omega_2)^2[\omega_3\omega_4 ((\vec k_1\cdot\vec k_2) - k_1 k_2)\right.\nonumber\\ &+\omega_1\omega_2 ((\vec k_3\cdot\vec k_4) - k_3 k_4)]\frac{1}{g^2}\nonumber\\
&-2(\omega_1 -\omega_3)^{2}[\omega_2\omega_4 ((\vec k_1\cdot\vec k_3) + k_1 k_3) + \omega_1\omega_3 ((\vec k_2\cdot\vec k_4) + k_2 k_4)]\frac{1}{g^2}\nonumber\\
&-2(\omega_1 -\omega_4)^{2}[\omega_2\omega_3 ((\vec k_1\cdot\vec k_4) + k_1 k_4) + \omega_1\omega_4 ((\vec k_2\cdot\vec k_3) + k_2 k_3)]\frac{1}{g^2}\nonumber\\
&+[(\vec k_1\cdot\vec k_2) + k_1 k_2][(\vec k_3\cdot\vec k_4) + k_3 k_4]\nonumber\\
& + [-(\vec k_1\cdot\vec k_3) + k_1 k_3][-(\vec k_2\cdot\vec k_4) + k_2 k_4]\\
&+[-(\vec k_1\cdot\vec k_4) + k_1 k_4][-(\vec k_2\cdot\vec k_3) + k_2 k_3]\nonumber\\
&+ 4(\omega_1 + \omega_2)^2\frac{[(\vec k_1\cdot\vec k_2) - k_1 k_2][(\vec k_3\cdot\vec k_4) - k_3 k_4]}{\omega^2_{\vec k_1 +\vec k_2} - (\omega_1 + \omega_2)^2}\nonumber\\
&+4(\omega_1 - \omega_3)^2\frac{[(\vec k_1\cdot\vec k_3) + k_1 k_3][(\vec k_2\cdot\vec k_4) + k_2 k_4]}{\omega^2_{\vec k_1 - \vec k_3} - (\omega_1 - \omega_3)^2}\nonumber\\
&\left. + 4(\omega_1 - \omega_4)^2\frac{[(\vec k_1\cdot\vec k_4) + k_1 k_4][(\vec k_2\cdot\vec k_3) + k_2 k_3]}{\omega^2_{\vec k_1-\vec k_4} - (\omega_1 - \omega_4)^2}\right\}.\label{T_tilde_1234_general}
\end{flalign}
Expression~\eqref{T_1234_general_PZ}  satisfies conditions $T_{\vec k_1 \vec k_2}^{\vec k_3 \vec k_4,{PRZ}}=T_{\vec k_2 \vec k_1}^{\vec k_3 \vec k_4,{PRZ}}=T_{\vec k_1 \vec k_2}^{\vec k_4 \vec k_3,{PRZ}}$ but lacks an important symmetry $T_{\vec k_1 \vec k_2}^{\vec k_3 \vec k_4} = T_{\vec k_3 \vec k_4}^{\vec k_1 \vec k_2}$ (analog of Hermitian symmetry in quantum mechanics). This symmetry is a consequence of the fact that system is Hamiltonian with a real Hamiltonian function. As a result, we need to perform an additional symmetrization:
\begin{equation}
\label{T_1234_general_KNPS}
T_{\vec k_1 \vec k_2}^{\vec k_3 \vec k_4} = \frac{1}{2}(T_{\vec k_1 \vec k_2}^{\vec k_3 \vec k_4,{PRZ}} + T_{\vec k_3 \vec k_4}^{\vec k_1 \vec k_2,{PRZ}}).
\end{equation}

\subsection{Supplemental Material: Part II.A. Wave system in  presence of condensate.}
We consider the following situation: two wave vectors ($\vec k_2$ and $\vec k_4$) are out of the condensate, and  two other wave vectors are in the condensate ($\vec k_1$ and $\vec k_3$). In accordance with derivation of diffusion equation in the main paper,the condensate wave vectors must be approximately on the same circle
%($n_{\vec k}= n_0 \delta(k-k_0)$), so 
$|\vec k_1|=|\vec k_3|=k_0$. 
%Because everything is isotropic with respect to an angle, one can choose $\vec k_2=\vec k$ 
Since the interaction coefficient is invariant with respect to the $k$-space rotations, we can
take $\vec k_2$ directed along the $x$-axis and measure the angles on the condensate circle with respect to this direction: $\vec k_1 = (k_0 \cos(\beta), k_0 \sin(\beta))^T$ and $\vec k_3 = (k_0 \cos(\alpha), k_0 \sin(\alpha))^T$, where $\beta = \angle (\vec k_2, \vec k_1)$ and $\alpha = \angle (\vec k_2, \vec k_3)$.

Resonant condition for the wave vectors is as follows,
\begin{equation}
\vec k_1 + \vec k_2 = \vec k_3 + \vec k_4,
\end{equation}
and for the frequencies -- 
\begin{equation}
\omega_1 + \omega_2 = \omega_3 + \omega_4.
\end{equation}
Because the condensate is a long-wave background, we can consider the case $k_0 \ll k_{2,4}$ where $k_i=|\vec k_i|$. So we have a small parameter $k_0/k_{2,4} \ll 1$. This is our major tool for simplification (reduction) of the matrix element. 
For the purposes of this paper, all we need is the case 
%$\vec k_2 = \vec k_4$, but even for the more general case $\vec k_2 \ne \vec k_4$ it can be shown (see the next section) that at least for the first few non-vanishing orders of expansion one can take 
$\vec k_2 = \vec k_4 = \vec k$.
As a result, to get the full matrix element, one needs to consider the following two functions, $\tilde T_{\vec k_1 \vec k}^{ \vec k_3 \vec k}$ and $\tilde T_{\vec k \, \vec k_1}^{ \vec k_3 \vec k}$.

\subsection{Supplemental Material: Part II.B. $\tilde T_{\vec k_1 \vec k}^{\vec k_3 \vec k}$ contribution.}
Let us begin with the $\tilde T_{\vec k_1\vec k}^{\vec k_3 \vec k}$:
\begin{flalign}
&\tilde T_{\vec k_1 \vec k}^{\vec k_3 \vec k} = -\frac{1}{16\pi^2}\frac{1}{(k_0 k)^{1/2}}\nonumber\\
&\times\left\{-12k_0^2 k^2 - 2(\omega_0 + \omega_k)^2[\omega_0\omega_k ((\vec k_1\cdot\vec k) - k_0 k)\right.\nonumber\\ &+\omega_0\omega_k ((\vec k_3\cdot\vec k) - k_0 k)]\frac{1}{g^2}\nonumber\\
&-2(0)^{2}[\omega_k^2 ((\vec k_1\cdot\vec k_3) + k_0^2) + \omega_0^2 ((2k^2)]\frac{1}{g^2}\nonumber\\
&-2(\omega_0 -\omega_k)^{2}[\omega_k\omega_0 ((\vec k_1\cdot\vec k) + k_0 k) + \omega_0\omega_k ((\vec k\cdot\vec k_3) + k k_0)]\frac{1}{g^2}\nonumber\\
&+[(\vec k_1\cdot\vec k) + k_0 k][(\vec k_3\cdot\vec k) + k_0 k]\nonumber\\
& + [-(\vec k_1\cdot\vec k_3) + k_0^2][0]\nonumber\\
&+[-(\vec k_1\cdot\vec k) + k_0 k][-(\vec k\cdot\vec k_3) + k k_0]\nonumber\\
&+ 4(\omega_0 + \omega_k)^2\frac{[(\vec k_1\cdot\vec k) - k_0 k][(\vec k_3\cdot\vec k) - k_0 k]}{\omega^2_{\vec k_1+\vec k} - (\omega_0 + \omega_k)^2}\nonumber\\
&+4
%(0)^2
(\omega_3 -\omega_1)^2 \frac{[(\vec k_1\cdot\vec k_3) + k_0^2][(2k^2)]}{\omega^2_{\vec k_1-\vec k_3} - 
%(0)^2
(\omega_3 -\omega_1)^2}\nonumber\\
&\left. + 4(\omega_0 - \omega_k)^2\frac{[(\vec k_1\cdot\vec k) + k_0 k][(\vec k\cdot\vec k_3) + k k_0]}{\omega^2_{\vec k_1-\vec k} - (\omega_0 - \omega_k)^2}\right\}.\label{T_tilde_1k3k} 
\end{flalign}
In the main text we showed that $|\omega_3 -\omega_1| = \mathcal{O}(|\vec q|) $ and, therefore
$(\omega_3 -\omega_1)^2 = \mathcal{O}(|\vec q|^2) $, whereas $\omega^2_{\vec k_1-\vec k_3} =
|\vec q|$. Therefore, the second fractional term is   $\mathcal{O}(|\vec q|^3) $ , so it can be neglected.

\begin{comment}
    
We have $\vec q = \vec k_1 - \vec k_3
%$ 
%taking into account that the choice of $\vec k_2=\vec k$ as described above (along the $x$-axis), angles $\alpha$ and $\beta$ are just the usual polar angles, so $\vec q 
= (k_0 (\cos\beta-\cos\alpha), k_0 (\sin\beta-\sin\alpha))^T$, so that
\begin{equation}
|\vec q| = k_0\sqrt{2(1-\cos(\beta-\alpha))},\label{abs_Delta_k}
\end{equation}
and
%can be taken as zero as if we take the limit 
for $\alpha\rightarrow\beta$, 
%according to~\eqref{abs_Delta_k}:
\begin{equation*}
|\Delta k|  \to k_0\sqrt{2}\frac{(\beta-\alpha)^2}{2},
\end{equation*}
so the term is always zero for any $\alpha\neq \beta$.
\end{comment}

Using the following expressions:
\begin{flalign}
(\omega_0 + \omega_k)^2 &= gk\left(1+2\sqrt{\frac{k_0}{k}} + \frac{k_0}{k}\right)\approx gk\left(1+2\sqrt{\frac{k_0}{k}}\right),\nonumber\\
(\omega_0 - \omega_k)^2 &= gk\left(1-2\sqrt{\frac{k_0}{k}} + \frac{k_0}{k}\right)\approx gk\left(1-2\sqrt{\frac{k_0}{k}}\right),\nonumber\\
\omega^2_{\vec k_1+\vec k} &= gk\sqrt{1+2\frac{k_0}{k}\cos\beta + \left(\frac{k_0}{k}\right)^2}\approx gk\left(1+\frac{k_0}{k}\cos\beta\right),\nonumber\\
\omega^2_{\vec k_1-\vec k} &= gk\sqrt{1-2\frac{k_0}{k}\cos\beta + \left(\frac{k_0}{k}\right)^2}\approx gk\left(1-\frac{k_0}{k}\cos\beta \right),\nonumber\\
\omega^2_{\vec k_3-\vec k} &= gk\sqrt{1-2\frac{k_0}{k}\cos\alpha + \left(\frac{k_0}{k}\right)^2}\approx gk\left(1-\frac{k_0}{k}\cos\alpha\right),
%\nonumber\\
\end{flalign}
%and supposing $\vec k_1 \neq \vec k_3$ (i.e. $\alpha \neq \beta$), 
one can get for the $1/g^2$-group of terms:
\begin{flalign}
&-2(\omega_0 + \omega_k)^2[\omega_0\omega_k ((\vec k_1\cdot\vec k) - k_0 k)+\omega_0\omega_k ((\vec k_3\cdot\vec k) - k_0 k)]\frac{1}{g^2}\nonumber\\
&-2(\omega_0 -\omega_k)^2[\omega_k\omega_0 ((\vec k_1\cdot\vec k) + k_0 k) + \omega_0\omega_k ((\vec k\cdot\vec k_3) + k k_0)]\frac{1}{g^2}=\nonumber\\
&=-2 k^2 k_0^2 \sqrt{\frac{k}{k_0}}\left(1+2\sqrt{\frac{k_0}{k}}+\frac{k_0}{k}\right)(\cos\beta+\cos\alpha-2)-\nonumber\\
&-2 k^2 k_0^2 \sqrt{\frac{k}{k_0}}\left(1-2\sqrt{\frac{k_0}{k}}+\frac{k_0}{k}\right)(\cos\beta+\cos\alpha+2)\approx\nonumber\\
&\approx-2 k^2 k_0^2 \sqrt{\frac{k}{k_0}}\left(1+2\sqrt{\frac{k_0}{k}}\right)(\cos\beta+\cos\alpha-2)-\nonumber\\
&-2 k^2 k_0^2 \sqrt{\frac{k}{k_0}}\left(1-2\sqrt{\frac{k_0}{k}}\right)(\cos\beta+\cos\alpha+2)=\nonumber\\
&=16 k^2 k_0^2 -4 k^2 k_0^2 \sqrt{\frac{k}{k_0}}(\cos\beta+\cos\alpha).\label{Over_g_terms}
\end{flalign}
For the terms in the square brackets we have:
\begin{flalign}
&[(\vec k_1\cdot\vec k) + k_0 k][(\vec k_3\cdot\vec k) + k_0 k]+[-(\vec k_1\cdot\vec k) + k_0 k][-(\vec k\cdot\vec k_3) + k k_0]=\nonumber\\
&=k^2 k_0^2 (\cos\beta + 1)(\cos\alpha + 1) + k^2 k_0^2 (1- \cos\beta)(1 - \cos\alpha)=\nonumber\\
&=2 k^2 k_0^2 + 2k^2 k_0^2\cos\beta \cos\alpha.\label{square_brakets_terms}
\end{flalign}
The first fractional term is:
\begin{flalign}
&4(\omega_0 + \omega_k)^2\frac{[(\vec k_1\cdot\vec k) - k_0 k][(\vec k_3\cdot\vec k) - k_0 k]}{\omega^2_{\vec k_1+\vec k} - (\omega_0 + \omega_k)^2}=\nonumber\\
&=4k\left(1+2\sqrt{\frac{k_0}{k}} + \frac{k_0}{k}\right)\frac{k^2 k_0^2(\cos\beta-1)(\cos\alpha-1)}{k\sqrt{1+2\frac{k_0}{k}\cos\beta + \left(\frac{k_0}{k}\right)^2} - k\left(1+2\sqrt{\frac{k_0}{k}} + \frac{k_0}{k}\right)}.\nonumber
\end{flalign}
Now one needs to use expansion of the square root in the denominator up to the $k_0/k$ terms:
\begin{equation}
\sqrt{1+2\frac{k_0}{k}\cos\beta + \left(\frac{k_0}{k}\right)^2}\approx 1 + \frac{k_0}{k}\cos\beta.
\end{equation}
\begin{flalign}
&4(\omega_0 + \omega_k)^2\frac{[(\vec k_1\cdot\vec k) - k_0 k][(\vec k_3\cdot\vec k) - k_0 k]}{\omega^2_{\vec k_1+\vec k} - (\omega_0 + \omega_k)^2}\approx\nonumber\\
&\approx 4k^2 k_0^2\left(1+2\sqrt{\frac{k_0}{k}} \right)\frac{(\cos\beta\cos\alpha+1 - (\cos\beta+\cos\alpha))}{- 2\sqrt{\frac{k_0}{k}} - \frac{k_0}{k}(1-\cos\beta)}\approx\nonumber\\
&\approx-2k^2 k_0^2\sqrt{\frac{k}{k_0}}\left(1+2\sqrt{\frac{k_0}{k}}\right)(1+\cos\beta\cos\alpha - (\cos\beta+\cos\alpha))\times\nonumber\\
&\times\left(1 - \frac{1}{2}\sqrt{\frac{k_0}{k}}(1-\cos\beta)\right),\label{first_fraction}
\end{flalign}
here we used expansion  up to $\sqrt{k_0/k}$ terms:
\begin{equation}
\frac{1}{1 + \frac{1}{2}\sqrt{\frac{k_0}{k}}(1-\cos\beta)} \approx  1 - \frac{1}{2}\sqrt{\frac{k_0}{k}}(1-\cos\beta).
\end{equation}
The last fractional term:
\begin{flalign}
&4(\omega_0 - \omega_k)^2\frac{[(\vec k_1\cdot\vec k) + k_0 k][(\vec k_3\cdot\vec k) + k_0 k]}{\omega^2_{\vec k_1-\vec k} - (\omega_0 - \omega_k)^2}=\nonumber\\
&=4k\left(1-2\sqrt{\frac{k_0}{k}} + \frac{k_0}{k}\right)\frac{k^2 k_0^2(\cos\beta+1)(\cos\alpha+1)}{k\sqrt{1-2\frac{k_0}{k}\cos\beta + \left(\frac{k_0}{k}\right)^2} - k\left(1-2\sqrt{\frac{k_0}{k}} + \frac{k_0}{k}\right)}.\nonumber
\end{flalign}
Now one needs to use expansion of the square root in the denominator up to the $k_0/k$ terms:
%($(k_0/k)^2$ are not necessary):
\begin{equation}
\sqrt{1-2\frac{k_0}{k}\cos\beta + \left(\frac{k_0}{k}\right)^2}\approx 1 - \frac{k_0}{k}\cos\beta.
\end{equation}
\begin{flalign}
&4(\omega_0 - \omega_k)^2\frac{[(\vec k_1\cdot\vec k) + k_0 k][(\vec k_3\cdot\vec k) + k_0 k]}{\omega^2_{\vec k_1-\vec k} - (\omega_0 - \omega_k)^2}\approx\nonumber\\
&\approx 4k^2 k_0^2\left(1-2\sqrt{\frac{k_0}{k}}\right)\frac{(\cos\beta\cos\alpha+1 + (\cos\beta+\cos\alpha))}{2\sqrt{\frac{k_0}{k}} - \frac{k_0}{k}(1+\cos\beta)}\approx\nonumber\\
&\approx2k^2 k_0^2\sqrt{\frac{k}{k_0}}\left(1-2\sqrt{\frac{k_0}{k}}\right)(1+\cos\beta\cos\alpha + (\cos\beta+\cos\alpha))\times\nonumber\\
&\times\left(1 + \frac{1}{2}\sqrt{\frac{k_0}{k}}(1+\cos\beta) \right),\label{last_fraction}
\end{flalign}
here we used expansion (up to $\sqrt{k_0/k}$ terms):
\begin{equation}
\frac{1}{1 - \frac{1}{2}\sqrt{\frac{k_0}{k}}(1+\cos\beta)} \approx  1 + \frac{1}{2}\sqrt{\frac{k_0}{k}}(1+\cos\beta).
\end{equation}
Let us factor out $k^2 k_0^2\sqrt{k/k_0}$ terms from both fractional terms~\eqref{first_fraction} and~\eqref{last_fraction}:
\begin{flalign}
&-2k^2 k_0^2\sqrt{\frac{k}{k_0}}(1+\cos\beta\cos\alpha - (\cos\beta+\cos\alpha))+2k^2 k_0^2\sqrt{\frac{k}{k_0}}\times\nonumber\\
&\times(1+\cos\beta\cos\alpha + (\cos\beta+\cos\alpha))=4k^2 k_0^2\sqrt{\frac{k}{k_0}}(\cos\beta+\cos\alpha).\label{fractional_terms_anomalous}
\end{flalign}
This term cancels exactly with corresponding term in~\eqref{Over_g_terms}. So, there will be no terms like that in the final expansion.

Now let us extract all $k^2 k_0^2$ terms from both fractional terms~\eqref{first_fraction} and~\eqref{last_fraction}:
\begin{flalign}
&-4k^2 k_0^2(1+\cos\beta\cos\alpha - (\cos\beta+\cos\alpha))+k^2 k_0^2(1+\cos\beta\cos\alpha - (\cos\beta+\cos\alpha))(1-\cos\beta)+\nonumber\\
&-4k^2 k_0^2(1+\cos\beta\cos\alpha + (\cos\beta+\cos\alpha))+k^2 k_0^2(1+\cos\beta\cos\alpha + (\cos\beta+\cos\alpha))(1+\cos\beta)=\nonumber\\
&=-6k^2 k_0^2 - 4k^2 k_0^2\cos\beta\cos\alpha + 2k^2 k_0^2\cos^2\beta,\label{fractional_terms_main}
\end{flalign}
here we have used
\begin{flalign}
&(1+\cos\beta\cos\alpha - (\cos\beta+\cos\alpha))(1-\cos\beta)+(1+\cos\beta\cos\alpha + (\cos\beta+\cos\alpha))(1+\cos\beta)=\nonumber\\
&=2+4\cos\beta\cos\alpha + 2\cos^2\beta.
\end{flalign}
Taking into account $-12k^2 k_0^2$ term at the beginning of the curly brackets in~\eqref{T_tilde_1k3k} and isotropic terms in~\eqref{Over_g_terms},\eqref{square_brakets_terms}, and~\eqref{fractional_terms_main} one can see that such terms cancel.

Angular dependent $k^2 k_0^2$-terms from~\eqref{square_brakets_terms} and~\eqref{fractional_terms_main} (pay attention, they are absent in~\eqref{Over_g_terms}) give:
\begin{equation}
-2k^2 k_0^2\cos\beta\cos\alpha + 2k^2 k_0^2\cos^2\beta = 2k^2 k_0^2\cos\beta(\cos\beta-\cos\alpha).\label{main_contribution_1}
\end{equation}
This term will vanish if $\cos\beta-\cos\alpha=0$. For every given $\beta$ there are only two values of $\alpha=\pm\beta$ when this is satisfied. For example, for a trivial process $T_{\vec k_1 \vec k_2}^{\vec k_1 \vec k_2}$ (in our case it corresponds to $\vec k_1 = \vec k_3$, which is prohibited) this term will be absent. But in general it is present and is the main contribution term.

As a result, for $\tilde T_{\vec k_1 \vec k}^{\vec k_3 \vec k}$ the first nonvanishing term is:
\begin{equation}
\label{tilde_T_1k3k}
\tilde T_{\vec k_1 \vec k}^{\vec k_3 \vec k} \approx -\frac{1}{16\pi^2}\frac{1}{(k_0 k)^{1/2}}2k^2 k_0^2\cos\beta(\cos\beta-\cos\alpha) = -\frac{(k k_0)^{3/2}}{16\pi^2} 2\cos\beta(\cos\beta-\cos\alpha).
\end{equation}

\subsection{Supplemental Material: Part II.C. $\tilde T_{\vec k \vec k_1}^{\vec k_3 \vec k}$ contribution.}
Let us recall, that $T_{\vec k_1 \vec k}^{\vec k_3 \vec k,PRZ} = (\tilde T_{\vec k_1 \vec k}^{\vec k_3 \vec k} + \tilde T_{\vec k \vec k_1}^{\vec k_3 \vec k})/2$, so
one needs to find all corresponding terms for $\tilde T_{\vec k \vec k_1}^{\vec k_3 \vec k}$ as well,
\begin{flalign}
&\tilde T_{\vec k \vec k_1}^{\vec k_3 \vec k} = -\frac{1}{16\pi^2}\frac{1}{(k_0 k)^{1/2}}\nonumber\\
&\times\left\{-12k_0^2 k^2 - 2(\omega_k + \omega_0)^2[\omega_0\omega_k ((\vec k\cdot\vec k_1) - k_0 k)\right.\nonumber\\ &+\omega_0\omega_k ((\vec k_3\cdot\vec k) - k_0 k)]\frac{1}{g^2}\nonumber\\
&-2(\omega_k -\omega_0)^{2}[\omega_0\omega_k ((\vec k\cdot\vec k_3) + k_0 k) + \omega_0\omega_k ((\vec k_1\cdot\vec k) + k k_0)]\frac{1}{g^2}-\nonumber\\
&-2(0)^{2}[\omega_0^2 (2k^2) + \omega_k^2 ((\vec k_1\cdot\vec k_3) + k_0^2)]\frac{1}{g^2}+\nonumber\\
&+[(\vec k\cdot\vec k_1) + k_0 k][(\vec k_3\cdot\vec k) + k_0 k]\nonumber\\
&+[-(\vec k\cdot\vec k_3) + k_0 k][-(\vec k_1\cdot\vec k) + k k_0]\\
& + [0][-(\vec k_1\cdot\vec k_3) + k_0^2]\nonumber\\
&+ 4(\omega_k + \omega_0)^2\frac{[(\vec k\cdot\vec k_1) - k_0 k][(\vec k_3\cdot\vec k) - k_0 k]}{\omega^2_{\vec k+\vec k_1} - (\omega_k + \omega_0)^2}\nonumber\\
&+ 4(\omega_k - \omega_0)^2\frac{[(\vec k\cdot\vec k_3) + k_0 k][(\vec k_1\cdot\vec k) + k k_0]}{\omega^2_{\vec k-\vec k_3} - (\omega_k - \omega_0)^2}\nonumber\\
&\left.+4(\omega_k-\omega_{k_4})^2\frac{[2k^2][(\vec k_1\cdot\vec k_3) + k_0^2]}{\omega^2_{\vec k - \vec k_4} - (\omega_k - \omega_{k_4})^2}\right\}.\label{T_tilde_k13k}
\end{flalign}
Here the last fractional term can be neglected for the similar reason as the one given in the paragraph after equation \eqref{T_tilde_1k3k}.
\begin{comment}
    
taken as zero, as $(\omega_k-\omega_{k_4})^2/k$ in the numerator and denominator is $\sim {k^2_0}/{k^2}$, while $\omega^2_{\vec k -\vec k_4}/k$ in the denominator is (here we use~\eqref{abs_Delta_k}):
\begin{equation}
\frac{\omega^2_{\vec k -\vec k_4}}{k} = \frac{|\vec k -\vec k_4|}{k} = \frac{|\vec q|}{k} = \frac{k_0}{k}\sqrt{2(1-\cos(\beta-\alpha))},
\end{equation}
meaning that leading term in the denominator is $\sim {k_0}/{k}$, so the whole fraction is $\sim k_0^2k^2\frac{k_0}{k}$ and negligible, because the first nonvanishing term is $\sim k_0^2 k^2$ (it is angular dependent). We neglect all further terms.
\end{comment}

As one can see, all leading order terms are the same as for $\tilde T_{\vec k_1 \vec k}^{\vec k_3 \vec k}$ with exception of the denominator of the second fractional term, where ($\omega^2_{\vec k_1-\vec k}$ is replaced by $\omega^2_{\vec k-\vec k_3}$). So let us work only with this term:
%(the second fractional term):
\begin{flalign}
&4(\omega_0 - \omega_k)^2\frac{[(\vec k_1\cdot\vec k) + k_0 k][(\vec k_3\cdot\vec k) + k_0 k]}{\omega^2_{\vec k-\vec k_3} - (\omega_0 - \omega_k)^2}=\nonumber\\
&=4k\left(1-2\sqrt{\frac{k_0}{k}} + \frac{k_0}{k}\right)\frac{k^2 k_0^2(\cos\beta+1)(\cos\alpha+1)}{k\sqrt{1-2\frac{k_0}{k}\cos\alpha + \left(\frac{k_0}{k}\right)^2} - k\left(1-2\sqrt{\frac{k_0}{k}} + \frac{k_0}{k}\right)}.\nonumber
\end{flalign}
Now one needs to use expansion of the square root in the denominator up to the $k_0/k$ terms:
%($(k_0/k)^2$ are not necessary):
\begin{equation}
\sqrt{1-2\frac{k_0}{k}\cos\alpha + \left(\frac{k_0}{k}\right)^2}\approx 1 - \frac{k_0}{k}\cos\alpha.
\end{equation}
\begin{flalign}
&4(\omega_0 - \omega_k)^2\frac{[(\vec k_1\cdot\vec k) + k_0 k][(\vec k_3\cdot\vec k) + k_0 k]}{\omega^2_{\vec k-\vec k_3} - (\omega_0 - \omega_k)^2}\approx\nonumber\\
&\approx 4k^2 k_0^2\left(1-2\sqrt{\frac{k_0}{k}} \right)\frac{(\cos\beta\cos\alpha+1 + (\cos\beta+\cos\alpha))}{2\sqrt{\frac{k_0}{k}} - \frac{k_0}{k}(1+\cos\alpha)}\approx\nonumber\\
&\approx2k^2 k_0^2\sqrt{\frac{k}{k_0}}\left(1-2\sqrt{\frac{k_0}{k}} \right)(1+\cos\beta\cos\alpha + (\cos\beta+\cos\alpha))\times\nonumber\\
&\times\left(1 + \frac{1}{2}\sqrt{\frac{k_0}{k}}(1+\cos\alpha) \right),\label{second_fraction}
\end{flalign}
here we used expansion (up to $\sqrt{k_0/k}$ terms):
\begin{equation}
\frac{1}{1 - \frac{1}{2}\sqrt{\frac{k_0}{k}}(1+\cos\alpha)} \approx  \left(1 + \frac{1}{2}\sqrt{\frac{k_0}{k}}(1+\cos\alpha) \right).
\end{equation}
Let us extract $k^2 k_0^2\sqrt{k/k_0}$ terms from both fractional terms~\eqref{first_fraction} and~\eqref{second_fraction}:
\begin{flalign}
&-2k^2 k_0^2\sqrt{\frac{k}{k_0}}(1+\cos\beta\cos\alpha - (\cos\beta+\cos\alpha))+2k^2 k_0^2\sqrt{\frac{k}{k_0}}\times\nonumber\\
&\times(1+\cos\beta\cos\alpha + (\cos\beta+\cos\alpha))=4k^2 k_0^2\sqrt{\frac{k}{k_0}}(\cos\beta+\cos\alpha).\label{fractional_terms_anomalous_2}
\end{flalign}
This term cancels exactly with corresponding term in~\eqref{Over_g_terms}. So, there will be NO terms like that in the final expansion.

Now let us extract all $k^2 k_0^2$ terms from both fractional terms~\eqref{first_fraction} and~\eqref{second_fraction}:
\begin{flalign}
&-4k^2 k_0^2(1+\cos\beta\cos\alpha - (\cos\beta+\cos\alpha))+k^2 k_0^2(1+\cos\beta\cos\alpha - (\cos\beta+\cos\alpha))(1-\cos\beta)+\nonumber\\
&-4k^2 k_0^2(1+\cos\beta\cos\alpha + (\cos\beta+\cos\alpha))+k^2 k_0^2(1+\cos\beta\cos\alpha + (\cos\beta+\cos\alpha))(1+\cos\alpha)=\nonumber\\
&=-6k^2 k_0^2 - 4k^2 k_0^2\cos\beta\cos\alpha + k^2 k_0^2\cos^2\beta + k^2 k_0^2\cos^2\alpha + k^2 k_0^2(\cos\alpha-\cos\beta)(\cos\beta\cos\alpha+1),\label{fractional_terms_main_2}
\end{flalign}
here we used:
\begin{flalign}
&(1+\cos\beta\cos\alpha - (\cos\beta+\cos\alpha))(1-\cos\beta)+(1+\cos\beta\cos\alpha + (\cos\beta+\cos\alpha))(1+\cos\alpha)=\nonumber\\
&=2+4\cos\beta\cos\alpha + \cos^2\beta + \cos^2\alpha + (\cos\alpha-\cos\beta)(\cos\beta\cos\alpha+1).
\end{flalign}
Taking into account $-12k^2 k_0^2$ term at the beginning of curly brackets of~\eqref{T_tilde_k13k} and isotropic terms in~\eqref{Over_g_terms},\eqref{square_brakets_terms}, and~\eqref{fractional_terms_main_2} one can see that such terms cancel.

Angular dependent $k^2 k_0^2$-terms from~\eqref{square_brakets_terms} and~\eqref{fractional_terms_main_2} (pay attention, they are absent in~\eqref{Over_g_terms}) give:
\begin{flalign}
&-2k^2 k_0^2\cos\beta\cos\alpha + k^2 k_0^2\cos^2\beta + k^2 k_0^2\cos^2\alpha + k^2 k_0^2(\cos\alpha-\cos\beta)(\cos\beta\cos\alpha+1) =\nonumber\\
&=k^2 k_0^2(\cos\alpha-\cos\beta)(\cos\alpha - \cos\beta + \cos\alpha\cos\beta + 1).\label{main_contribution_2}
\end{flalign}
This term will vanish if $\cos\beta-\cos\alpha=0$. For every given $\beta$ there are only two values of $\alpha=\pm\beta$ when this is satisfied. For example, for a trivial process $T_{\vec k_1 \vec k_2, \vec k_1 \vec k_2}$ (in our case it corresponds to $\vec k_1 = \vec k_3$, which is prohibited) this term will be absent. But in general it is present and is the main contribution term.

Finally, for $\tilde T_{\vec k \vec k_1}^{\vec k_3 \vec k}$ we get the first nonvanishing term:
\begin{flalign}
\tilde T_{\vec k \vec k_1}^{\vec k_3 \vec k} &= \approx -\frac{1}{16\pi^2}\frac{1}{(k_0 k)^{1/2}}k^2 k_0^2(\cos\alpha-\cos\beta)(\cos\alpha - \cos\beta + \cos\alpha\cos\beta + 1) =\nonumber\\
&=-\frac{(k k_0)^{3/2}}{16\pi^2} (\cos\alpha-\cos\beta)(\cos\alpha - \cos\beta + \cos\alpha\cos\beta + 1).\label{tilde_T_k13k}
\end{flalign}

\subsection{Supplemental Material: Part II.D. The first nonvanishing term of matrix element.}
If one takes arithmetic mean value of~\eqref{tilde_T_1k3k} and~\eqref{tilde_T_k13k} we get the main term in $T_{\vec k_1\vec k}^{\vec k_3\vec k,PRZ}$:
\begin{equation}
\label{preliminary_main_contribution}
T_{\vec k_1\vec k}^{\vec k_3\vec k,PRZ}=\frac{\tilde T_{\vec k_1 \vec k}^{\vec k_3 \vec k} + \tilde T_{\vec k \vec k_1}^{\vec k_3 \vec k}}{2}\approx -\frac{(k k_0)^{3/2}}{32\pi^2} (\cos\alpha-\cos\beta)(\cos\beta\cos\alpha+1+\cos\alpha-3\cos\beta).
\end{equation}
It should be noted, that the expression is not symmetric with respect to exchange of $\alpha$ and $\beta$. This is exactly the consequence of lack of symmetry with respect to exchange of pairs of vectors (the lower and the upper pair) $T_{\vec k_1\vec k_2}^{\vec k_3\vec k_4} = T_{\vec k_3\vec k_4}^{\vec k_1\vec k_2}$, which for our case means $T_{\vec k_1\vec k}^{\vec k_3\vec k} = T_{\vec k_3\vec k}^{\vec k_1\vec k}$, or $T(\vec k, k_0, \alpha,\beta) = T(\vec k, k_0,\beta, \alpha)$. The expression~\eqref{T_1234_general_KNPS} after substitution of~\eqref{preliminary_main_contribution} takes the form:
\begin{equation}
\label{total_main_contribution_matrix_element}
T_{\vec k_1\vec k}^{\vec k_3\vec k} = \frac{T(\vec k, k_0, \alpha,\beta) + T(\vec k, k_0,\beta, \alpha)}{2} = -\frac{(k k_0)^{3/2}}{16\pi^2}(\cos\alpha -\cos\beta)^2.
\end{equation}
In the waves kinetic equations we use square of the~\eqref{total_main_contribution_matrix_element} and one gets:
\begin{equation}
\label{squared_main_contribution_matrix_element}
|T_{\vec k_1\vec k}^{\vec k_3\vec k}|^2 = \frac{(k k_0)^{3}}{256\pi^4}(\cos\alpha -\cos\beta)^4.
\end{equation}
For a flux in an angularly symmetric case one needs to compute:
\begin{equation}
D(k) = 2\pi k \int\limits_{0, k_1 \ll k}^{+\infty} n^2_{k_1} k_1^{9/2}\int\limits_{0}^{2\pi}\int\limits_{0}^{2\pi}|T_{\vec k_1\vec k}^{\vec k_3\vec k}|^2(\cos\beta-\cos\alpha)^2\D\alpha\D\beta\D k_1
\end{equation}
We can integrate it (average by angles) over the interval $[0,2\pi)$ for both $\alpha$ and $\beta$, as there is no dependence on the angle (situation is isotropic with respect to polar angles), then one gets: 
\begin{equation}
\label{angle_averaged_squared_main_contribution_matrix_element}
\int\limits_{0}^{2\pi}\int\limits_{0}^{2\pi} |T_{\vec k_1\vec k}^{\vec k_3\vec k}|^2(\cos\beta-\cos\alpha)^2\D\alpha\D\beta = \frac{(k k_1)^{3}}{256\pi^4}25\pi^2= \frac{25 k_1^{3}}{256\pi^2}k^3.
\end{equation}

\subsection{Supplemental Material: Part II.E. Symbolic computations of the next order term of matrix element expansion.}
We used Maxima Computer Algebra System\cite{Maxima} (specifically {\tt wxMaxima for Linux}) in order to compute the next order term in the expansion of the matrix element $T_{\vec k_1\vec k}^{\vec k_3\vec k}$ and to check all previous calculations. In order to simplify expressions, we factored out
$$
-\frac{1}{16\pi^2}\frac{(k k_0)^2}{(k k_0)^{1/2}}=-\frac{(k k_0)^{3/2}}{16\pi^2}
$$
from all the terms and had to consider expansion of
$$
\frac{k_4}{k}=\frac{|\vec k + \vec\Delta k|}{k}.
$$
The small parameter $k_0/k$ is denoted by {\tt x}.

Here is the code:
\begin{lstlisting}[breaklines]
/* In order to allow for next order terms, we take into account relative difference between k2=k and k4. */;
k4_over_k(x,beta,alpha):=taylor(sqrt(1+2*x*(cos(beta)-cos(alpha)+2*x^2*(1-cos(beta-alpha)))),x,0,2);

/* For bravity we denote \tilde T with just T. */

/* This is part II.B of Supplemental Material. */
T_1234_first_term(x,beta,alpha):=-12*k4_over_k(x,beta,alpha);

T_1234_over_g_2_term_1(x,beta,alpha):=taylor(-2*(sqrt(x)+1)^2*(sqrt(k4_over_k(x,beta,alpha))*(cos(beta)-1)+(cos(alpha)+x*(cos(beta-alpha)-1)-k4_over_k(x,beta,alpha))),x,0,2)/sqrt(x);

T_1234_over_g_2_term_2(x,beta,alpha):=-0;

T_1234_over_g_2_term_3(x,beta,alpha):=taylor(-2*((sqrt(x)-sqrt(k4_over_k(x,beta,alpha)))^2)*((cos(beta)+x*(1-cos(beta-alpha))+k4_over_k(x,beta,alpha))+sqrt(k4_over_k(x,beta,alpha))*(cos(alpha)+1)),x,0,2)/sqrt(x);

T_1234_over_g_2_terms(x,beta,alpha):=T_1234_over_g_2_term_1(x,beta,alpha)+T_1234_over_g_2_term_2(x,beta,alpha)+T_1234_over_g_2_term_3(x,beta,alpha);

T_1234_square_brackets_term_1(x,beta,alpha):=taylor((cos(beta)+1)*(cos(alpha)+x*(cos(beta-alpha)-1)+k4_over_k(x,beta,alpha)),x,0,2);

T_1234_square_brackets_term_2(x,beta,alpha):=taylor((-cos(beta-alpha)+1)*(-(1+x*(cos(beta)-cos(alpha)))+k4_over_k(x,beta,alpha)),x,0,2);

T_1234_square_brackets_term_3(x,beta,alpha):=taylor((-cos(alpha)+1)*(-(cos(beta)+x*(1-cos(beta-alpha)))+k4_over_k(x,beta,alpha)),x,0,2);

T_1234_square_brackets_terms(x,beta,alpha):=T_1234_square_brackets_term_1(x,beta,alpha)+T_1234_square_brackets_term_2(x,beta,alpha)+T_1234_square_brackets_term_3(x,beta,alpha);

T_1234_fractional_term_1(x,beta,alpha):=taylor(4*(1+sqrt(x))^2*(cos(beta)-1)*(cos(alpha)+x*(cos(beta-alpha)-1)-k4_over_k(x,beta,alpha))/(sqrt(1+2*x*cos(beta)+x^2)-(1+sqrt(x))^2),x,0,2);

T_1234_fractional_term_2(x,beta,alpha):=0;

T_1234_fractional_term_3(x,beta,alpha):=taylor(4*(sqrt(x)-sqrt(k4_over_k(x,beta,alpha)))^2*(cos(alpha)+1)*(cos(beta)+x*(1-cos(beta-alpha))+k4_over_k(x,beta,alpha))/(sqrt((k4_over_k(x,beta,alpha))^2-2*x*(cos(beta)+x*(1-cos(beta-alpha)))+x^2)-(sqrt(x)-sqrt(k4_over_k(x,beta,alpha)))^2),x,0,2);

T_1234_fractional_terms(x,beta,alpha):=T_1234_fractional_term_1(x,beta,alpha)+T_1234_fractional_term_2(x,beta,alpha)+T_1234_fractional_term_3(x,beta,alpha);

T_1234(x,beta,alpha):=T_1234_first_term(x,beta,alpha)+T_1234_over_g_2_terms(x,beta,alpha)+T_1234_square_brackets_terms(x,beta,alpha)+T_1234_fractional_terms(x,beta,alpha);

/* In order to see expansion, we evaluate the \tilde T as in the end of part II.B of Supplemental Material, Eq. (26). */
T_1234(x,beta,alpha);

/* This is part II.C of Supplemental Material. */
T_2134_over_g_2_term_2(x,beta,alpha):=taylor(-2*(1-sqrt(x))^2*(sqrt(k4_over_k(x,beta,alpha))*(cos(alpha)+1)+(cos(beta)+x*(1-cos(beta-alpha))+k4_over_k(x,beta,alpha))),x,0,2)/sqrt(x);

T_2134_over_g_2_term_3(x,beta,alpha):=taylor(-2*(1-sqrt(k4_over_k(x,beta,alpha)))^2*((1+x*(cos(beta)-cos(alpha))+k4_over_k(x,beta,alpha))+x*(cos(beta-alpha)+1)),x,0,2)/x;

/* The first term divided by g^2 is the same as in part II.B. */

T_2134_over_g_2_terms(x,beta,alpha):=T_1234_over_g_2_term_1(x,beta,alpha)+T_2134_over_g_2_term_2(x,beta,alpha)+T_2134_over_g_2_term_3(x,beta,alpha);

T_2134_square_brackets_term_2(x,beta,alpha):=taylor((-cos(alpha)+1)*(-(cos(beta)+x*(1-cos(beta-alpha)))+k4_over_k(x,beta,alpha)),x,0,2);

T_2134_square_brackets_term_3(x,beta,alpha):=taylor((-1-x*(cos(beta)-cos(alpha))+k4_over_k(x,beta,alpha))*(-cos(beta-alpha)+1),x,0,2);

T_2134_square_brackets_terms(x,beta,alpha):=T_1234_square_brackets_term_1(x,beta,alpha)+T_2134_square_brackets_term_2(x,beta,alpha)+T_2134_square_brackets_term_3(x,beta,alpha);

/* The first fractional term is the same as in part II.B. */

T_2134_fractional_term_2(x,beta,alpha):=taylor(4*(1-sqrt(x))^2*(cos(alpha)+1)*(cos(beta)+x*(1-cos(beta-alpha))+k4_over_k(x,beta,alpha))/(sqrt(1-2*x*cos(alpha)+x^2)-(1-sqrt(x))^2),x,0,2);

T_2134_fractional_term_3(x,beta,alpha):=0;

T_2134_fractional_terms(x,beta,alpha):=T_1234_fractional_term_1(x,beta,alpha)+T_2134_fractional_term_2(x,beta,alpha);

/* The firs term of the whole expression is the same as in part II.B. */

T_2134(x,beta,alpha):=T_1234_first_term(x,beta,alpha)+T_2134_over_g_2_terms(x,beta,alpha)+T_2134_square_brackets_terms(x,beta,alpha)+T_2134_fractional_terms(x,beta,alpha);

/* In order to see expansion, we evaluate the \tilde T as in the end of part II.C of Supplemental Material, Eq. (36). */
T_2134(x,beta,alpha);

/* This is part II.D of Supplemental Material, Eq. (37). */
T_PRZ(x,beta,alpha):=(T_1234(x,beta,alpha)+T_2134(x,beta,alpha))/2;

/* Evaluation of Eq. (37). */
T_PRZ(x,beta,alpha);

/* This is Eq. (38),  part II.D of Supplemental Material. */
T_full(x,beta,alpha):=(T_PRZ(x,beta,alpha)+T_PRZ(x,alpha,beta))/2;

/* Evaluation of final expression of matrix element, as in Eq. (38) of Supplemental Material. */
T_full(x,beta,alpha);
\end{lstlisting}

One can see, that the expansion of the $T_{\vec k_1\vec k}^{\vec k_3\vec k}$ up to the next order is:
\begin{equation*}
T_{\vec k_1\vec k}^{\vec k_3\vec k} \approx -\frac{(k k_0)^{3/2}}{16\pi^2}\left\{(\cos\alpha -\cos\beta)^2 + \frac{(\cos\alpha + \cos\beta)\left[(\cos\alpha-\cos\beta)^2 -4 -4\cos(\alpha-\beta)\right]}{2}\sqrt{\frac{k_0}{k}}\right\}.
\end{equation*}
\newpage

\section{Supplemental Material: Part III. Derivation of Diffusion Equation from the Zakharov Equation}
Below, we will derive a WKB description for water gravity waves on background of a large-scale condensate wave component followed by assuming that the waves are weak and random, and obtaining the same spectral diffusion equation as the one derived in the main text. 
The aim of this excersize is to show that the limit of weak nonlinearity commutes with the scale separation limit. Our approach will be similar in spirit as the one used for a three-wave MHD system in \cite{NAZARENKO2001646} except here, for the first time, we apply it to a four-wave system.

We start from the Zakharov equation of surface gravity waves in standard form:
\begin{equation}
    i\frac{\partial b_{\vec k}}{\partial t} = \omega_{\vec k} b_{\vec k} + \int T_{\vec k\vec k_1}^{\vec k_2\vec k_3} b^*_{\vec 1} b_{\vec 2} b_{\vec 3} \delta (\vec k+\vec k_1-\vec k_2-\vec k_3) d \vec k_1 d\vec k_2 d\vec k_3,
    \label{ZAK}
\end{equation}
with $\omega_{\vec k} = k^{1/2}$, $k=|\vec k|$ the dispersion relation, and $b_{\vec k}$ the canonical variable obtained from a near-identity Lee transform of variable $a_{\vec k}$ defined in the main paper. We consider the condensate condition, where \eqref{ZAK} is dominated by non-local interactions of long and short waves, and we take 
\begin{equation}
    \vec k, \vec k_2: \text{large wavenumbers}; \ \  \vec k_1, \vec k_3: \text{small wavenumbers}, \ \ 
    \label{scasep}
\end{equation}
with scale separation, e.g., $k\gg k_1$.
%$|\vec k| \gg |\vec k_1|$. 
Accordingly \eqref{ZAK} reduces to
\begin{equation}
    i\frac{\partial b_{\vec k}}{\partial t} = k^{1/2}b_{\vec k} + 2\int d\vec k_2 b_{\vec 2} \int_{1,3 \text{ small}} T_{\vec k\vec k_1}^{\vec k_2\vec k_3} b^*_{\vec 1} b_{\vec 3} \delta (\vec k+\vec k_1-\vec k_2-\vec k_3) d \vec k_1 d\vec k_3,
    \label{ZAK_SEP}
\end{equation}
where the factor ``2'' comes from the fact that we can exchange $\vec k_2$ and $\vec k_3$ in \eqref{scasep}. We further define
\begin{equation}
    A(\vec k_2,\vec k)=\int_{1,3 \text{ small}} T_{\vec k\vec k_1}^{\vec k_2\vec k_3} b^*_{\vec 1} b_{\vec 3} \delta (\vec k+\vec k_1-\vec k_2-\vec k_3) d \vec k_1 d\vec k_3,
\end{equation}
and 
\begin{equation}
    B(\vec k_2,\vec k)=k^{1/2} \delta(\vec k-\vec k_2) +  2A(\vec k_2,\vec k),
\end{equation}
so that \eqref{ZAK_SEP} can be written as
\begin{equation}
    i\frac{\partial b_{\vec k}}{\partial t} = \int B(\vec k_2,\vec k) b_{\vec 2} d\vec k_2.
    \label{ZAK_SEP_B}
\end{equation}
We note that $A(\vec k_2,\vec k)$ satisfies the symmetry property
\begin{equation}
\begin{split}
    A^*(\vec k_2,\vec k) & =\int_{1,3 \text{ small}}    T_{\vec k\vec k_1}^{\vec k_2\vec k_3} b_{\vec 1} b^*_{\vec 3} \delta (\vec k+\vec k_1-\vec k_2-\vec k_3) d \vec k_1 d\vec k_3 \\
    & = \int_{1,3 \text{ small}}    T_{\vec k\vec k_3}^{\vec k_2\vec k_1} b_{\vec 3} b^*_{\vec 1} \delta (\vec k+\vec k_3-\vec k_2-\vec k_1) d \vec k_3 d\vec k_1 \\
    & = A(\vec k,\vec k_2),
\end{split}
\end{equation}
where in the last equality we have used the symmetry property of the interaction coefficient $T_{\vec k\vec k_3}^{\vec k_2\vec k_1}=T_{\vec k_2\vec k_1}^{\vec k\vec k_3}$ that is required for \eqref{ZAK} to be a Hamiltonian system with real Hamiltonian (see discussion in part I). In addition, we have
\begin{equation}
    B^*(\vec k_2,\vec k) = B(\vec k,\vec k_2).
\end{equation}

The derivation of the WKB (Liouville) equation requires the definition of wave action $n(\vec k,\vec x)$ (neglecting time dependence in definition for simplicity). One of the approaches for this is to use the Wigner transform, formulated here as the inverse Fourier transform of correlation in wavenumber space (c.f. \cite{DYACHENKO1992330}):
\begin{equation}
    n(\vec k,\vec x)=\int e^{i\vec q\cdot\vec x} \langle b_{\vec k+\vec q/2}b^*_{\vec k-\vec q/2} \rangle d\vec q,
    \label{WIGNER}
\end{equation}
with the angle bracket representing ensemble average. It follows that
\begin{equation}
    \frac{\partial n}{\partial t} = \int e^{i\vec q\cdot\vec x} \partial_t \langle b_{\vec k+\vec q/2}b^*_{\vec k-\vec q/2} \rangle d\vec q,
\end{equation}
where
\begin{equation}
\begin{split}
    \partial_t b_{\vec k+\vec q/2} = -i \int B(\vec k_2,\vec k+\vec q/2) b_{\vec 2} d\vec k_2, \\
    \partial_t b^*_{\vec k-\vec q/2} = i \int B^*(\vec k_2,\vec k-\vec q/2) b^*_{\vec 2} d\vec k_2, 
\end{split}
\end{equation}
so that
\begin{equation}
\begin{split}
    \frac{\partial n}{\partial t} & =i \int d\vec k_2 d\vec q e^{i\vec q\cdot\vec x} \big[B^*(\vec k_2,\vec k-\vec q/2) \langle b^*_{\vec 2} b_{\vec k+\vec q/2} \rangle - B(\vec k_2,\vec k+\vec q/2) \langle b_{\vec 2} b^*_{\vec k-\vec q/2} \rangle \big] \\
    & = i \int d\vec k_2 d\vec q e^{i\vec q\cdot\vec x} \big[B(\vec k-\vec q/2, \vec k_2) \langle b_{\vec k+\vec q/2} b^*_{\vec 2} \rangle - B(\vec k_2,\vec k+\vec q/2) \langle b_{\vec 2} b^*_{\vec k-\vec q/2} \rangle\big].
\end{split}
\end{equation}

We further introduce change of variables $\vec k_2 = \vec k + \vec q''/2 - \vec q'/2$ and $\vec q = \vec q'' + \vec q'$, then
\begin{equation}
\begin{split}
    \frac{\partial n}{\partial t} = i \int d\vec q' d\vec q'' e^{i(\vec q''+\vec q')\cdot\vec x} \big[ B(\vec k-\vec q''/2-\vec q'/2, \vec k+\vec q''/2-\vec q'/2) \langle b_{\vec k+\vec q''/2+\vec q'/2} b^*_{\vec k+\vec q''/2-\vec q'/2} \rangle \\ - B(\vec k+\vec q''/2-\vec q'/2,\vec k+\vec q''/2+\vec q'/2) \langle b_{\vec k+\vec q''/2-\vec q'/2} b^*_{\vec k-\vec q''/2-\vec q'/2} \rangle\big].
    \label{nt_q}
\end{split}
\end{equation}

We define a condensate-modulated frequency 
\begin{equation}
    \Omega(\vec k,\vec x)=\int e^{i\vec q\cdot\vec x} B(\vec k-\vec q/2, \vec k+\vec q/2) d\vec q,
    \label{MODFREQ}
\end{equation}

Inverting \eqref{WIGNER} and \eqref{MODFREQ}, we obtain
\begin{equation}
   \langle b_{\vec k+\vec q/2}b^*_{\vec k-\vec q/2} \rangle = \frac{1}{4\pi^2} \int e^{-i\vec q\cdot\vec x} n(\vec k,\vec x) d\vec x,
   \label{WIGINV}
\end{equation}
\begin{equation}
   B(\vec k-\vec q/2, \vec k+\vec q/2) = \frac{1}{4\pi^2} \int e^{-i\vec q\cdot\vec x} \Omega(\vec k,\vec x) d\vec x,
   \label{MFEINV}
\end{equation}
Substituting \eqref{WIGINV} and \eqref{MFEINV} into \eqref{nt_q} gives
\begin{equation}
\begin{split}
    \frac{\partial n}{\partial t} & = i \int d\vec q' d\vec q'' e^{i(\vec q''+\vec q')\cdot\vec x} \Big[ \frac{1}{4\pi^2} \int e^{-i\vec q''\cdot\vec x''}\Omega(\vec k-\vec q'/2,\vec x'') d\vec x'' \frac{1}{4\pi^2} \int e^{-i\vec q'\cdot\vec x'}n(\vec k+\vec q''/2,\vec x') d\vec x' \\ & \ \ \ \ \ \ \ \ \ \ \ - \frac{1}{4\pi^2} \int e^{-i\vec q'\cdot\vec x'}\Omega(\vec k+\vec q''/2,\vec x') d\vec x' \frac{1}{(4\pi)^2} \int e^{-i\vec q''\cdot\vec x''}n(\vec k-\vec q'/2,\vec x'') d\vec x'' \Big].
    \label{nt_complex} \\ & = \frac{i}{(4\pi^2)^2} \int d\vec q' d\vec q'' d\vec x' d\vec x'' e^{i\vec q''\cdot(\vec x-\vec x'')+i\vec q'\cdot(\vec x-\vec x')} \\ & \ \ \ \ \ \ \ \ \ \ \ \ \ \ \ \ \ \ \ \ \  \Big[\Omega(\vec k-\vec q'/2,\vec x'') n(\vec k+\vec q''/2,\vec x') - \Omega(\vec k+\vec q''/2,\vec x')n(\vec k-\vec q'/2,\vec x'') \Big]
\end{split}
\end{equation}
Applying the symmetry $\vec q',\vec x' \leftrightarrow \vec q'',\vec x''$ to the first term of the above equation, we obtain
\begin{equation}
\begin{split}
    \frac{\partial n}{\partial t} = & \frac{i}{(4\pi^2)^2} \int d\vec q' d\vec q'' d\vec x' d\vec x'' e^{i\vec q''\cdot(\vec x-\vec x'')+i\vec q'\cdot(\vec x-\vec x')} \\ & \ \ \ \ \ \ \ \ \ \ \ \ \ \ \ \ \ \ \ \ \  \Big[\Omega(\vec k-\vec q''/2,\vec x') n(\vec k+\vec q'/2,\vec x'') - \Omega(\vec k+\vec q''/2,\vec x')n(\vec k-\vec q'/2,\vec x'') \Big].
\end{split}
\end{equation}
The term in the square bracket of the above equation can be Taylor expanded as (considering $\vec q'$, $\vec q''$ as small quantities relative to $\vec k$)
\begin{equation}
    [...]=-\vec q''\cdot \frac{\partial \Omega(\vec k,\vec x')}{\partial \vec k} n(\vec k,\vec x'') + \vec q'\cdot \frac{\partial n(\vec k,\vec x'')}{\partial \vec k}\Omega(\vec k,\vec x')
    \label{sbra}
\end{equation}
To continue, we take the first term in \eqref{sbra} as an example. We can combine the factor $-i\vec q''$ with one exponential to obtain $(\partial/\partial \vec x'')e^{i\vec q''\cdot(\vec x-\vec x'')}$, and perform integration by parts to move the $\vec x''$ derivative to $n(\vec k,\vec x'')$. Applying this procedure to both terms in \eqref{sbra} and using the relation $\int d\vec q e^{i\vec q\cdot \vec x}/4\pi^2=\delta(\vec x)$ results in 
\begin{equation}
    \frac{\partial n}{\partial t} + \frac{\partial \Omega(\vec k,\vec x)}{\partial \vec k} \cdot  \frac{\partial n(\vec k,\vec x)}{\partial \vec x} - \frac{\partial \Omega(\vec k,\vec x)}{\partial \vec x} \cdot  \frac{\partial n(\vec k,\vec x)}{\partial \vec k} = 0.
    \label{WKB}
\end{equation}
This is the WKB equation, which can also be written in the form of a Liouville equation:
\begin{equation}
    \frac{\partial n}{\partial t} + \{ H,n \}=0,
    \label{liouville}
\end{equation}
where $\{  \}$ represents the Poisson bracket, and
\begin{equation}
\begin{split}
    H & =\Omega(\vec k,\vec x) = \int d\vec q e^{i\vec q\cdot \vec x} B(\vec k-\vec q/2, \vec k+\vec q/2) \\ & = \int d\vec q e^{i\vec q\cdot \vec x} \big[ |\vec k+\vec q/2|^{1/2} \delta(\vec q) + 2A(\vec k-\vec q/2, \vec k+\vec q/2) \big] \\ & = |\vec k|^{1/2} + 2\int d\vec q e^{i\vec q\cdot \vec x} \int_{1,3 \text{ small}}   T_{\vec k+\vec q/2,\vec k_1}^{\vec k-\vec q/2,\vec k_3} b^*_{\vec 1} b_{\vec 3} \delta (\vec q+\vec k_1-\vec k_3) d \vec k_1 d\vec k_3,
    \label{H_Omega}
\end{split}
\end{equation}

Our next goal is to derive the diffusion equation from \eqref{WKB} or \eqref{liouville}. We consider $b^*_{\vec 1} b_{\vec 3} \sim O(\epsilon)$, and re-write \eqref{H_Omega} as
\begin{equation}
    \Omega(\vec k,\vec x) = \omega(\vec k) + \epsilon \Omega_0 (\vec k,\vec x),
    \label{purt1}
\end{equation}
where $\omega(\vec k)=|\vec k|^{1/2}$ is the inherent frequency and $\epsilon \Omega_0 (\vec k,\vec x)$ is the remaining (small) condensate-modulated component.

We also rewrite \eqref{WKB} in standard ray-tracing form:
\begin{equation}
    \frac{\partial n}{\partial t} + \dot{\vec x} \cdot  \frac{\partial n(\vec k,\vec x)}{\partial \vec x} + \dot{\vec k} \cdot  \frac{\partial n(\vec k,\vec x)}{\partial \vec k} = 0,
    \label{RAYTR}
\end{equation}
with ray equations
\begin{equation}
    \dot{\vec x} = \frac{\partial \Omega(\vec k,\vec x)}{\partial \vec k}, \ \ \dot{\vec k} = -\frac{\partial \Omega(\vec k,\vec x)}{\partial \vec x} = -\epsilon \frac{\partial \Omega_0(\vec k,\vec x)}{\partial \vec x}.
    \label{ray}
\end{equation}

Since $\nabla \cdot \dot{\vec x} = -\nabla_{\vec k} \cdot \dot{\vec k}$, \eqref{ray} can be formulated as
\begin{equation}
    \frac{\partial n}{\partial t} + \nabla \cdot (\dot{\vec x}n) + \nabla_{\vec k} \cdot (\dot{\vec k}n) = 0. 
    \label{RAYTR2}
\end{equation}

Let
\begin{equation}
n(\vec k,\vec x,t) = \bar{n}(\vec k,t_d) + \tilde{n}(\vec k,\vec x,t),
\end{equation}
where $\bar{n}$ is the spatially homogeneous part and $\tilde{n}$ the ``wiggles'' due to the large-scale condensate modulation. $t_d=\epsilon^2 t$ is a slow diffusion time which will be characterized in the final diffusion equation.

We consider randomized large-scale condensate motion with averaging operator $E[\cdot]$ that can be considered as ensemble average or spatial average $\int \cdot d\vec x/L^2$ with large $L$. Then it follows
\begin{equation}
\bar{n}(\vec k,t_d) = E[n(\vec k,\vec x,t)].
\end{equation}
and (using \eqref{RAYTR2} and \eqref{ray})
\begin{equation}
    \partial_t \bar{n} = -E[\nabla_{\vec k} \cdot (\dot{\vec k} n)] = \epsilon \nabla_{\vec k} \cdot \big( E[\frac{\partial \Omega_0}{\partial \vec x}(\bar{n} + \tilde{n})] \big) = \epsilon \nabla_{\vec k} \cdot \big( E[\frac{\partial \Omega_0}{\partial \vec x} \tilde{n}] \big),
    \label{diff0}
\end{equation}
where the last equality comes from the fact that $\bar{n}$ is independent of $\vec x$ so that the average over the large-scale condensate motion (for the first term) gives zero. 

We next expand the wave action along a ray:
\begin{equation}
\begin{split}
    n(\vec k,\vec x,t) & = n\Big(\vec k-\int_{t-T}^t \dot{\vec k}(t')dt', \vec x-\int_{t-T}^t \dot{\vec x}(t')dt', t-T \Big) \\
    & = \bar{n}(\vec k-\int_{t-T}^t \dot{\vec k}(t')dt',t_d - \epsilon^2 T) + \tilde{n}\Big(\vec k-\int_{t-T}^t \dot{\vec k}(t')dt', \vec x-\int_{t-T}^t \dot{\vec x}(t')dt', t-T \Big) \\ & = \bar{n}(\vec k,t_d) - \int_{t-T}^t \dot{\vec k}(t')dt' \cdot \nabla_{\vec k} \bar{n}(\vec k,t_d) - \epsilon^2 T \partial_t \bar{n}(\vec k,t_d) \\ & \ \ \ \ \ \ +  \tilde{n}\Big(\vec k-\int_{t-T}^t \dot{\vec k}(t')dt', \vec x-\int_{t-T}^t \dot{\vec x}(t')dt', t-T \Big).
    \label{rayexp}
\end{split}
\end{equation}

Keeping $O(1)$ terms in \eqref{rayexp} (we note that this is consistent with the later choice of time scale $T$), we obtain
\begin{equation}
    \tilde{n}(\vec k,\vec x,t) = - \int_{t-T}^t \dot{\vec k}(t')dt' \cdot \nabla_{\vec k} \bar{n}(\vec k,t_d) + \tilde{n}\Big(\vec k-\int_{t-T}^t \dot{\vec k}(t')dt', \vec x-\int_{t-T}^t \dot{\vec x}(t')dt', t-T \Big).
    \label{rayresult}
\end{equation}

We choose $T$ large enough so that the second term in \eqref{rayresult} is de-correlated with the large scale motion (but still smaller compared to the diffusion time scale), i.e., $T\sim O(1/\epsilon)$, then
\begin{equation}
    E\Big[\frac{\partial \Omega_0}{\partial \vec x} \tilde{n}\Big(\vec k-\int_{t-T}^t \dot{\vec k}(t')dt', \vec x-\int_{t-T}^t \dot{\vec x}(t')dt', t-T \Big) \Big] = 0.
    \label{avezero}
\end{equation}
Substituting \eqref{rayresult} to \eqref{diff0}, and considering \eqref{avezero}, we obtain
\begin{equation}
\begin{split}
    \partial_t \bar{n} & = - \epsilon \nabla_{\vec k} \cdot \Big( E\big[ \frac{\partial \Omega_0}{\partial \vec x} \big( \int_{t-T}^t \dot{\vec k}(t')dt' \cdot \nabla_{\vec k} \bar{n}(\vec k,t_d) \big)   \big]   \Big) \\ & = \epsilon^2 \nabla_{\vec k} \cdot \Big( E\big[ \frac{\partial \Omega_0}{\partial \vec x} \big( \int_{t-T}^t \partial_{\vec x} \Omega_0 (\vec k + \epsilon \frac{\partial \Omega_0}{\partial \vec x}(t-t'), \vec x- \frac{\partial \Omega}{\partial \vec k}(t-t'),t')dt'  \cdot \nabla_{\vec k} \bar{n}(\vec k,t_d) \big)   \big]   \Big) \\ & = \epsilon^2 \nabla_{\vec k} \cdot \Big( E\big[ \frac{\partial \Omega_0}{\partial \vec x} \big( \int_{t-T}^t \partial_{\vec x} \Omega_0 (\vec k, \vec x- \frac{\partial \omega}{\partial \vec k}(t-t'),t')dt'  \cdot \nabla_{\vec k} \bar{n}(\vec k,t_d) \big)   \big]   \Big),
    \label{diff_vec}
\end{split}
\end{equation}
where in the last equality we neglect the higher-order ($\epsilon^3$) terms. Equation \eqref{diff_vec} is in fact in the form of a diffusion equation. To see this, we write it in Einstein notation, and in the meanwhile absorb $\epsilon^2$ into $t$ on the LHS to form $t_d$, neglect the overbar on $n$ and subscript $d$ of $t$. This gives the final diffusion equation
\begin{equation}
    \frac{\partial n}{\partial t} = \partial_{k_i} (D_{ij} \partial_{k_j}n),
\end{equation}
with
\begin{equation}
    D_{ij} = E\Big[ \partial_{{x}_i} \Omega_0 \int_{t-T}^t \partial_{{x}_j}\Omega_0\big( \vec k, \vec x- \frac{\partial \omega}{\partial \vec k}(t-t'),t' \big) dt' \Big],
    \label{diffcoef}
\end{equation}
where (with a bit abuse of notation) $k_{i,j}$ represents the $i,j$th component of vector $\vec k$. 

We next perform detailed analysis on the diffusion coefficient \eqref{diffcoef} to show that it is equivalent to the result derived in the main paper from the wave kinetic equation. Under a change of variable $t'=t+s$, \eqref{diffcoef} reads
\begin{equation}
    D_{ij} = E\Big[ \partial_{{x}_i} \Omega_0 \int_{-T}^0 \partial_{{x}_j}\Omega_0\big( \vec k, \vec x + \frac{\partial \omega}{\partial \vec k}s,t+s \big) ds \Big]
    \label{diffcoef2}
\end{equation}
where 
\begin{equation}
    \partial_{{x}_i}\Omega_0 = 2i \int d\vec q q_i e^{i\vec q\cdot \vec x} \int_{1,3 \text{ small}} d \vec k_1 d\vec k_3 T_{\vec k+\vec q/2,\vec k_1}^{\vec k-\vec q/2,\vec k_3} b^*_{\vec 1} b_{\vec 3} \delta (\vec q+\vec k_1-\vec k_3) 
    \label{partialOme}
\end{equation}
\begin{equation}
    \Omega_0 \big( \vec k, \vec x + \frac{\partial \omega}{\partial \vec k}s,t+s \big) = 2 \int d\vec q' e^{i\vec q'\cdot (\vec x+\partial_{\vec k}\omega s)} \int_{2,4 \text{ small}} d \vec k_2 d\vec k_4 T_{\vec k+\vec q/2,\vec k_2}^{\vec k-\vec q/2,\vec k_4} b^*_{\vec 2}(t+s) b_{\vec 4}(t+s) \delta (\vec q'+\vec k_2-\vec k_4) 
\end{equation}
\begin{equation}
\begin{split}
    & \partial_{{x}_j}\Omega_0 \big( \vec k, \vec x + \frac{\partial \omega}{\partial \vec k}s,t+s \big) \\ & = 2i \int d\vec q' q_j'e^{i\vec q'\cdot (\vec x+\partial_{\vec k}\omega s)} \int_{2,4 \text{ small}} d \vec k_2 d\vec k_4 T_{\vec k+\vec q/2,\vec k_2}^{\vec k-\vec q/2,\vec k_4} b^*_{\vec 2}(t+s) b_{\vec 4}(t+s) \delta (\vec q'+\vec k_2-\vec k_4) 
\end{split}
\end{equation}
Assuming linear dynamics for the large-scale condensate motion, we have
\begin{equation}
    b^*_{\vec 2}(t+s) = b^*_{\vec 2}(t) e^{i\omega_2 s}, \ \ b_{\vec 4}(t+s) = b_{\vec 4}(t) e^{-i\omega_4 s}.
    \label{linearlarge}
\end{equation}
Substituting \eqref{partialOme}-\eqref{linearlarge} to \eqref{diffcoef2}, we obtain
\begin{equation}
\begin{split}
    D_{ij} & = E\Big[ -4 \int d\vec qd\vec q' q_i q_j' e^{i(\vec q'+\vec q) \cdot \vec x} \int_{1,2,3,4 \text{ small}} d \vec k_1 d\vec k_2 d \vec k_3 d\vec k_4 \\ & T_{\vec k+\vec q/2,\vec k_1}^{\vec k-\vec q/2,\vec k_3} T_{\vec k+\vec q/2,\vec k_2}^{\vec k-\vec q/2,\vec k_4} b^*_{\vec 1} b^*_{\vec 2} b_{\vec 3} b_{\vec 4} \int_{-T}^0 e^{i(\vec q' \cdot \partial_{\vec k}\omega +\omega_2 - \omega_4)s} ds \delta (\vec q+\vec k_1-\vec k_3) \delta (\vec q'+\vec k_2-\vec k_4)  \Big]
    \label{diffcoef3}
\end{split}
\end{equation}
For $T$ large enough, the $s$-integral becomes $\pi \delta (\vec q' \cdot \partial_{\vec k}\omega +\omega_2 - \omega_4)$ (which is half of the $-\infty$ to $\infty$ integration). The expectation operator can be distributed to $e^{i(\vec q'+\vec q) \cdot \vec x}$ and $b^*_{\vec 1} b^*_{\vec 2} b_{\vec 3} b_{\vec 4}$ due to independence. For the former, we need $\vec q'=-\vec q$ to keep \eqref{diffcoef3} to be nonzero. For the latter, Wick's rule can be applied with $E[b^*_{\vec 1} b^*_{\vec 2} b_{\vec 3} b_{\vec 4}] = n_1n_3\delta_{\vec 1 \vec 4}\delta_{\vec 2 \vec 3}$ (note that $\delta_{\vec 1 \vec 3}\delta_{\vec 2 \vec 4}$ is not possible due to other delta constraints in \eqref{diffcoef3}). Considering all above simplifications, we arrive at
\begin{equation}
    D_{ij} = 4\pi \int d\vec q q_i q_j \int_{1,3 \text{ small}} d \vec k_1 d \vec k_3 |T_{\vec k+\vec q/2,\vec k_1}^{\vec k-\vec q/2,\vec k_3}|^2 n_1n_3  \delta (\vec q \cdot \partial_{\vec k}\omega -\omega_3 + \omega_1) \delta (\vec q+\vec k_1-\vec k_3).
    \label{diff_final}
\end{equation}
In \eqref{diff_final}, we can change $\vec q$ into $-\vec q$ in the integrand using symmetry of the integration. Then, under the condensate condition, for which $q\ll k$ (so that Taylor expansion of the interaction coefficient leads to higher-order terms of $O(q^3)$ in the integrand which can be neglected), we see that \eqref{diff_final} is consistent with diffusion coefficient in the main paper. One can continue to make the isotropic assumption and solve for the stationary solution from here.
\newpage

\section{Supplemental Material: Part IV. Limits on small parameter $k_0/k$ or where our theory is applicable?}
Let us consider situation with given vector $\vec k$. Then the resonant conditions can be written as:
\begin{flalign}
\vec k_1 + \vec k = \vec k_3 + \vec k + \vec q,\\
\sqrt{k_1} + \sqrt{k} = \sqrt{k_3} + (|\vec k + \vec q|^2)^{1/4},
\end{flalign}
where $\vec k_1$ and $\vec k_3$ are vectors in condensate and $\vec q=\vec k_1 - \vec k_3$.
\begin{equation}
|\vec k + \vec q|^2 = k^2 + 2\vec k\cdot\vec q + |\vec q|^2 \approx  k^2 + 2k(k_1\cos\alpha - k_3\cos\beta).
\end{equation}
After expansion up to the first order in $k_{1,3}/k$ one gets:
\begin{flalign}
\sqrt{k_1} + \sqrt{k} = \sqrt{k_3} + \sqrt{k}\left[1 + \frac{1}{2}\left(\frac{k_1}{k}\cos\alpha - \frac{k_3}{k}\cos\beta\right) + \mathcal{O}\left(\frac{|\vec q|^2}{k^2}\right)\right],\label{sqrt_expansion_1}\\
2\frac{\sqrt{k_1} - \sqrt{k_3}}{\sqrt{k}}\approx\frac{k_1}{k}\cos\alpha - \frac{k_3}{k}\cos\beta.
\end{flalign}
The latter estimate implies that in resonant quartets $|k_3-k_1| \ll k_1, k_3 \approx k_0$. Indeed,  assuming {\em a priori} that this is true we have
%If we introduce a radius $k_0$ in the middle of the condensate ring, the following estimate for relatively narrow ($\epsilon/k_0\ll 1$) will work:
\begin{flalign}
\sqrt{k_1} - \sqrt{k_3} \approx 
%\left.\frac{d\omega_k}{d k}\right|_{k=k_0}(k_1-k_3) = \frac{1}{2}\frac{k_1 - k_3}{\sqrt{k_0}},\\
\frac{k_1 - k_3}{2\sqrt{k_0}}
\end{flalign}
which confirms the above strong inequality.
Then also
\begin{equation}
   \frac{k_1 - k_3}{\sqrt{kk_0}} \approx\frac{k_1}{k}\cos\alpha - \frac{k_3}{k}\cos\beta
   \approx \frac{k_0}{k}(\cos\alpha - \cos\beta).\label{alpha_beta_initial}
\end{equation}
so
%If we consider $\cos\beta$ to be a given value (we first integrate by $\alpha$ for every $\beta$), then in order to cover all possible values for $\cos\alpha$ we get a limit for $k$. For simplicity, let us consider case $\cos\beta=0$ and recall $k_1/k \approx k_0/k$:
\begin{flalign}
%\frac{k_1 - k_3}{\sqrt{k k_0}}\approx\frac{k_1}{k}\cos\alpha,\\
\cos\alpha -\cos \beta \approx \sqrt{\frac{k}{k_0}}\frac{k_1 - k_3}{k_0}.
\end{flalign}
For example, in numerical simulations of~\cite{Korotkevich2022arXiv} in the case of highest steepness $\mu\approx 0.135$ one can estimate $k_0\approx 8-9$, width of the ring (decay by the factor of $e$ from maximum) is  $\varepsilon \approx 4$. It means that we have an equation:
\begin{equation}
|\cos\alpha -\cos \beta| < \sqrt{\frac{k}{k_0}}\frac{\varepsilon}{k_0} \approx
0.5 \sqrt{\frac{k}{k_0}}.
\label{ring}
\end{equation}
So if $k/k_0\ge 16$ the RHS is greater than 2 and we can cover all possible values of $\cos\alpha - \cos \beta$, meaning that limits of integration are $\alpha,\beta\in[0;2\pi)$. 
%But this condition ($k_0/k\le 1/4$) has to be satisfied as we require $k_0/k\ll 1$ in all of our computations. 
%No \textcolor{blue}{Sasha: This condition is satisfied in your simulations?} 
Also, for any relative width of the condensate ring $\varepsilon/k_0$ we can limit ourselves with $k/k_0$ large enough to cover all possible values of $\alpha$ and $\beta$: $\alpha, \beta \in[0;2\pi)$. Indeed, here is the approximate relation for the relative width of the condensate ring $\varepsilon/k_0$  and the relative distance  $k_0/k$ (recall, that $|\cos\alpha -  \cos \beta|$ has to take all possible values from 0 to 2):
\begin{equation}
\frac{\varepsilon/k_0}{\sqrt{k_0/k}}\ge 2.
\end{equation}
In the other words, going far enough in $k$ (making the denominator small enough) we can make the values in the left hand side of the latter relation  reach the necessary threshold, after which one can integrate over the whole range $\alpha,\beta\in[0;2\pi)$. 

%\bibliographystyle{apsrev4-2}
%\bibliography{surfacewaves}
\end{document}